\documentclass[a4paper,11pt]{article}
\pdfoutput=1

\usepackage{scalerel}
\usepackage{jcappub}
\usepackage[T1]{fontenc}
\usepackage[utf8]{inputenc}
\usepackage{mathrsfs}
\usepackage{amsmath,amssymb}
\usepackage{mathrsfs}
\usepackage{xcolor}
\usepackage{slashed}
\usepackage[normalem]{ulem}
\usepackage{graphicx}
\usepackage{amssymb}
\usepackage{diagbox}

\newcommand{\dd}[1]{\mathrm{d}#1}
\newcommand{\Neff}{N_{\rm eff}}
\newcommand{\zfin}{z_{\rm fin}}

\title{\boldmath Towards a precision calculation of  the effective number of neutrinos $\Neff$ in the Standard Model I: The QED equation of state
}
\author[a]{Jack~J.~Bennett,}
\author[b]{Gilles~Buldgen,}
\author[b]{Marco~Drewes}
\author[a]{and Yvonne~Y.~Y.~Wong}

\affiliation[a]{School of Physics, The University of New South Wales, Sydney NSW 2052, Australia}
\affiliation[b]{Centre for Cosmology, Particle Physics and Phenomenology, Universit\'{e} catholique de Louvain, Louvain-la-Neuve B-1348, Belgium}

\emailAdd{j.j.bennett@unsw.edu.au, gilles.buldgen@uclouvain.be, marco.drewes@uclouvain.be, yvonne.y.wong@unsw.edu.au}

\date{\today}

\abstract{We revisit and quantify in this work several aspects of Standard Model physics at finite temperature that drive the theoretical value of the cosmological parameter, the effective number of neutrinos $N_{\rm eff}$, away from~3 in the early universe.  Our chief focus is  finite-temperature corrections to the equation of state of the QED plasma in the vicinity of neutrino decoupling at $T \sim 1$~MeV, where $T$ is the photon temperature.  Working in the instantaneous decoupling approximation,  we recover at ${\cal O}(e^2)$, where $e$ is the elementary electric charge,  the well-established correction of $\delta N_{\rm eff}^{(2)} \simeq 0.010$ across a range of plausible neutrino decoupling temperatures, in contrast to an erroneous claim in the recent literature which found twice as large an effect.   At ${\cal O}(e^3)$ we find a new and significant correction of  $\delta N_{\rm eff}^{(3)} \simeq -0.001$ that has so far not been accounted for in any precision neutrino decoupling calculation of~$N_{\rm eff}$, significant because this correction is in fact larger than---or at least comparable to---the change in $N_{\rm eff}$ induced between including and excluding neutrino oscillations in the transport modelling.  In addition to the QED equation of state, we make a first pass at quantifying finite-temperature QED corrections to the weak interaction rates that directly affect the neutrino decoupling process, and find in this connection that the ${\cal O}(e^2)$ thermal electron mass correction induces a change of  $\delta N_{\rm eff}^{m_{\rm th}}   \lesssim 10^{-4}$.
A complete assessment of the various effects considered in this work on the final value of $N_{\rm eff}$ will necessitate an account of neutrino energy transport beyond the instantaneous decoupling approximation.  However, relative to $N_{\rm eff}  = 3.044$ obtained in the most recent such calculation, we expect the new effects found in this work to lower the number to $N_{\rm eff} = 3.043$.
}

\begin{document}
\begin{flushright}
{\tt CP3-19-51}
\end{flushright}

\maketitle

\section{Introduction}

The concordance flat $\Lambda$CDM paradigm of cosmology has enjoyed remarkable success at explaining the structure of the universe on the largest scales. In this worldview,  the universe's large-scale evolution history can be captured by six parameters: the cold dark matter density~$\omega_{c}$, the baryon density~$\omega_{b}$, the reduced Hubble parameter~$h$, the amplitude and spectral index of the primordial curvature power spectrum~$A_{s}$ and~$n_{s}$, and the optical depth to reionisation~$\tau$.  With the advent of space-based cosmic microwave background (CMB) anisotropy probes such as the WMAP~\cite{Hinshaw:2012aka} and Planck missions~\cite{Aghanim:2018eyx}, these parameters have even been determined to better than 1\% precision; cosmology as a precision science has come of age.

Hidden beneath the spectacular precision of these results, however, is one crucial theoretical input, namely, the universe's energy density in Standard Model (SM) neutrinos relative to photons in the post-$e^{\pm}$-annihilation era (i.e., at temperatures $T \lesssim 0.5$~MeV).  Conventionally parameterised as the ``effective number of neutrinos''~$N_{\rm eff}$, this energy density ratio imprints on  CMB observables in several ways degenerate with the phenomenology of the concordance $\Lambda$CDM parameters~\cite{Hou:2011ec,Abazajian:2012ys}.  Such degeneracies immediately imply that how well $N_{\rm eff}$ is known theoretically must impact directly on the accuracy and precision with which we can infer $\Lambda$CDM parameters from observations, with further ramifications for the search  for beyond-the-SM light relics (e.g.,~\cite{Hannestad:2015tea,Abazajian:2019oqj}) and interactions (e.g.,~\cite{Diacoumis:2018nbq}).

Within the confines of the SM, the expected theoretical value of $N_{\rm eff}$ is~3---corresponding to three generations of left-handed neutrinos---plus percent-level corrections due to neutrino energy transport~\cite{Hannestad:1995rs,Dolgov:1997mb,Dolgov:1998sf,Esposito:2000hi,Mangano:2005cc,Escudero:2018mvt}  and finite-temperature quantum electrodynamics (FTQED)~\cite{Dicus:1982bz,Heckler:1994tv,Lopez:1998vk,Mangano:2001iu}.
Historically, estimates of these corrections---both individually and in combination---have yielded a range of values from $\delta N_{\rm eff} = 0.011$ to $0.052$~(see table VI of~\cite{Grohs2016} for a summary).  Improved numerical modelling of out-of-equilibrium energy transport at neutrino decoupling in the past 15 years, however, has narrowed the range to $\delta N_{\rm eff}= 0.044  \to 0.052$~\cite{Mangano:2005cc,deSalas,Birrell:2014uka,Grohs2016,Gariazzo:2019gyi}, with the 2006 result of $N_{\rm eff} = 3.046$~\cite{Mangano:2005cc}---which includes neutrino oscillations in the transport modelling---being most commonly quoted in the literature, and which was revised in 2019 to $N_{\rm eff} = 3.044$ to better than per-mille precision~\cite{Gariazzo:2019gyi}.

While presently too small to be of impact on parameter inference for the current generation of cosmological observations,%
\footnote{The current best observational constraint on $N_{\rm eff}$ is  $N_{\rm eff}=2.99^{+0.34}_{-0.33}$~(95\% C.I.)~\cite{Aghanim:2018eyx}, derived from the Planck TT+TE+EE+lowE+lensing+BAO data combination for a 7-parameter vanilla $\Lambda$CDM+$N_{\rm eff}$ model.}
 theoretical corrections and uncertainties of these magnitudes will begin to account for a sizeable fraction of the error budget---or even become measurable---as the parameter sensitivities of the next generation of experiments approach the sub-percent region.  The CMB-S4 experiment, for example,  is expected to improve the $1\sigma$ sensitivity to $N_{\rm eff}$ to $\sigma(N_{\rm eff}) \simeq 0.02 \to 0.03$~\cite{Abazajian:2016yjj}.  Planning has already begun in earnest for the construction of CMB-S4 to begin in as early as 2021~\cite{Abazajian:2019eic}. The time is therefore ripe for a closer scrutiny of the precision calculation of the theoretical $N_{\rm eff}$ in the context of the SM, and to beat down the theoretical/computational uncertainty of this crucial cosmological parameter to beyond the fourth significant digit.
 
In this first instalment of a series of papers  in which we revisit the $N_{\rm eff}$ calculation, we focus on FTQED corrections to the equation of state of the QED plasma in the vicinity of neutrino decoupling~($T \sim 1$~MeV).  The computation of FTQED corrections  in the context of $N_{\rm eff}$ has a long history, with the first investigation dating to 1982~\cite{Dicus:1982bz}.  
The dominant effect is a small departure of the QED equation of state  from the ideal gas limit, which is usually understood heuristically as a consequence of a self-energy-induced modification to the
dispersion relations of the electron/positron and the photon, i.e., through interactions, these particles pick up ``thermal masses" as they propagate in the QED plasma~\cite{Heckler:1994tv}.

Currently, the most widely used computational procedure to accounts for these  equation of state effects in the context of $N_{\rm eff}$ 
is that documented in~\cite{Mangano:2001iu,Mangano:2005cc}, which follows the same heuristic thermal mass arguments of~\cite{Heckler:1994tv} to ${\cal O}(e^{2})$, where $e$ is the elementary electric charge.  In the absence of transport corrections, 
the procedure yields at leading-order  $\delta N_{\rm eff} \simeq 0.01$.  Interestingly, this result was recently challenged in~\cite{Grohs2016}, where it was found that the non-ideal gas behaviour of the QED plasma contributes as much as  $\delta N_{\rm eff} \simeq 0.02$.  Indeed, this unusually large FTQED correction appears to be the main driver behind their likewise irregular final value of $\Neff=3.052$~\cite{Grohs2016} (which includes both transport and FTQED corrections) relative to the canonical $N_{\rm eff} = 3.044$~\cite{Gariazzo:2019gyi}.  

Our first and most urgent goal in this work, therefore, is to pinpoint the exact sources of discrepancy in the computational procedure of the QED equation of state that have led to these divergent results.
Along the way we also quantify the impact of a number of subdominant  ${\cal O}(e^{2})$ to ${\cal O}(e^{4})$ contributions on $N_{\rm eff}$  that have not yet been seriously considered in previous calculations.  In particular, we shall show that the ${\cal O}(e^{3})$ contribution produces a subleading deviation in $N_{\rm eff}$ that is  larger than---or, at least, comparable to---the difference induced between  including and excluding neutrino oscillations in the modelling of out-of-equilibrium energy transport at neutrino decoupling~\cite{Mangano:2005cc}, and is hence a 
necessary input if the SM  $N_{\rm eff}$ is to be computed to four-digit significance.

 The paper is organised as follows. We begin  in section~\ref{sec:physical} with a description of the relevant physical system and the standard approximations, and outline the estimation of $N_{\rm eff}$ from entropy conservation arguments as well as from solution of the continuity equation.  Section~\ref{sec:nevercoupled} discusses the main correction to $N_{\rm eff}$, which results from dropping what we shall call the ``ultra-relativistic approximation''.  We introduce finite-temperature effects to the QED equation of state in section~\ref{FTQED_intro}, from which we compute corrections to $N_{\rm eff}$ to ${\cal O}(e^{4})$, and pinpoint the error that had led to the discrepant result of~\cite{Grohs2016}.  In section~\ref{real_time_FTQFT} we use real-time finite-temperature field theory to compute the neutrino damping rate, which we use in conjunction with the Hubble expansion rate to estimate the neutrino decoupling temperature.  We conclude in section~\ref{sec:conclusions}.  Two appendices detail, respectively, the correspondence of the damping rate from finite-temperature field theory to the Boltzmann collision integral from kinetic theory, and the computation of the neutrino damping rate at leading order.


\section{The physical system}
\label{sec:physical}

Consider an epoch in the early universe when the photon bath attains a temperature of $T \sim 10$~MeV.%
\footnote{Unless explicitly labelled with subscripts ${\nu}, e$, or $\gamma$, all thermodynamic quantities (e.g., $T, \rho, P, s$, etc.) pertain to the combined system of the photon bath and the electron/positron population that is, in the timeframe of interest, always in thermal equilibrium with it.}
Within the standard hot big bang model coupled with the SM of particle physics, the universe's energy density at this time  is expected to be dominated by an ultra-relativistic QED plasma of photons and electrons/positrons, plus three generations of SM neutrinos and their antiparticles, all kept in thermal equilibrium and hence at the same temperature by the particles' weak and/or electromagnetic interactions with one another. See, e.g.,~\cite{lesgourgues2013neutrino} for a review.  For our purposes, the system can be taken to be (i) homogeneous and isotropic, and (ii) $CP$-symmetric with a vanishing chemical potential~$\mu$.

With the expansion of space come the dilution and adiabatic cooling of this primordial plasma and hence a decline in the particle interaction rates. For the system at hand, two major events ensue: 
\begin{itemize}
 \item {\bf Neutrino decoupling}:  At $T \sim 1$~MeV,  the interaction rate between the weakly-interacting neutrino sector and the electron/positron fluid drops below the Hubble expansion rate~$H$.  Hereafter, the neutrinos and the QED plasma lose thermal contact with one another, meaning that the temperatures of the two sectors, $T_{\nu}$ and $T$, are no longer bound to be the same.
  
 \item  {\boldmath $e^\pm$} {\bf annihilation}: At $T \sim 0.5$~MeV, the QED plasma cools to temperatures below the electron rest mass $m_{e} = 0.511$~MeV.  Here, kinematics favour the annihilation of electron/positron pairs into photons, leading to a net transfer of entropy from the electron/positron population to the photon sector.
 \end{itemize}

For our problem at hand, the main consequence of these two events is that the neutrinos emerge from them at the low temperature end ($T/m_{e} \to 0$) a little cooler than the photons. 
In an idealised scenario wherein 
\begin{enumerate}
\setlength{\itemsep}{-0.0em} 
\item All particle species behave as ideal gases ({\bf ideal gas approximation}), 
\item  The neutrino decoupling process is localised at $T= T_{\nu} = T_{d}$ ({\bf instantaneous decoupling approximation}), i.e., the neutrino and QED sectors transit from a state of tight thermal contact to a state of zero thermal contact at the instant $T_{d}$, where $T_{d}$ is a nominal neutrino decoupling temperature, and 
\item The electron/positron sector is fully ultra-relativistic at the time of neutrino decoupling, i.e., $T_d/m_e \to \infty$ ({\bf ultra-relativistic approximation}),
\end{enumerate}
the relative coolness of the neutrinos to the photons can be quantified by a simple temperature relation, $T_{\nu} = (4/11)^{1/3} T$,
based upon entropy conservation arguments (e.g.,~\cite{lesgourgues2013neutrino}).
It then follows that, post-$e^{\pm}$-annihilation, the energy density carried in the neutrino sector $\left.\rho_{\nu} \right|_{T/m_{e} \to 0}$  can be related to the photon energy density $\left.\rho_{\gamma} \right|_{T/m_{e} \to 0}$ via,
\begin{equation}
\label{Neff_def}
\left. \rho_\nu  \right|_{T/m_{e} \to 0}=\frac{7}{8} \Big(\frac{4}{11}\Big)^{4/3} N_{\rm eff}\  \left. \rho_\gamma  \right|_{T/m_{e} \to 0},
\end{equation}
where, in the idealised scenario, the $N_{\rm eff}$ parameter value is by definition 3, corresponding to three generations of SM neutrinos.

Clearly, relaxing any one of the above three approximations will induce a departure from the idealised $\Neff=3$.   We shall investigate in sections~\ref{sec:nevercoupled} and \ref{FTQED_intro}  respectively  the effects of relaxing the ultra-relativistic  and the ideal gas approximations, both of which are analytically tractable using the methods outlined below in sections~\ref{sec:entropy} and~\ref{sec:ode}, 
provided the assumption of instantaneous decoupling is maintained.  Relaxing the instantaneous decoupling approximation as well generally necessitates that we solve a system of quantum kinetic equations~\cite{Sigl:1992fn,McKellar:1992ja}
 numerically; this calculation will be presented in a subsequent publication.


\subsection{Estimating $N_{\rm eff}$ from entropy conservation arguments}
\label{sec:entropy}

Consider a time {\it after} neutrino decoupling at which the scale factor is $a$.  Here, frequent collisions within the QED plasma are able to keep it in a state of quasi-static thermal equilibrium over the entire timeframe of interest.  Likewise, the SM neutrino sector, though now technically composed of non-colliding particles, also continues to maintain its ideal-gas equilibrium phase space distribution to a good approximation, if the decoupling had been instantaneous.  Thus, the entropies in a comoving volume residing in the two sectors can be taken to be separately conserved (e.g.,~\cite{lesgourgues2013neutrino}), i.e., 
\begin{align}
s (a_{1}) a_{1}^{3} &= s (a_{2}) a_{2}^{3}, \label{eq:segammaconservation}\\
s_{\nu} (a_{1}) a_{1}^{3} &= s_{\nu} (a_{2}) a_{2}^{3}, \label{eq:snuconservation}
\end{align}
where $s \equiv s_{\gamma}+s_{e}$ and $s_{\nu}$ denote, respectively, the entropy density of the QED plasma and of the neutrino sector.

We take the scale factor $a_{1}$  to correspond to the epoch of neutrino decoupling such that $T_{\nu}(a_{1}) = T(a_{1})= T_{d}$.  The entropy densities are then given by
\begin{align}
s(a_{1})  & = \left. s^{(0)} \right|_{T_{d}/m_{e} \to \infty} + \delta s, \label{eq:segammaR1} \\
s_{\nu} (a_{1})& = 3 \times  \frac{7}{8} \frac{2 \pi^{2}}{45} g_{\nu} T_{\nu}^{3} (a_{1}), \label{eq:snu1}
\end{align}
where  
\begin{equation}
 \left. s^{(0)}\right|_{T_{d}/m_{e} \to \infty} = \frac{2 \pi^{2}}{45} \left(g_{\gamma} +\frac{7}{8} g_{e}\right) T_d^{3} 
\end{equation}
is the QED entropy density under the dual approximations of an ideal gas (represented by superscript ``(0)'') that is also ultra-relativistic ($T_{d}/m_{e} \to \infty$),  $\delta s$ is the change in the entropy density when either or both of these assumptions are relaxed, and $g_{\gamma}=g_{\nu}=2$ and $g_{e}=4$ denote the numbers of internal degrees of freedom of the various particle species.

We take $a_{2}$ to denote a time well {\it after} the nominal $e^{\pm}$ annihilation epoch, i.e., where $T/m_{e} \to 0$, when the electron/positron population is fully non-relativistic and significantly depleted relative to the photon population.  In this limit, the ideal gas approximation holds for both the QED and neutrino sectors, and 
\begin{align}
s (a_{2})  &= \frac{2 \pi^{2}}{45} g_{\gamma} T^{3} (a_{2}), \label{eq:segammaR2} \\
s_{\nu} (a_{2})& = 3 \times  \frac{7}{8} \frac{2 \pi^{2}}{45} g_{\nu} T_{\nu}^{3} (a_{2}) \label{eq:snu2}
\end{align}
are the relevant entropy densities.

Equating~(\ref{eq:snu1}) and (\ref{eq:snu2}) via equation~(\ref{eq:snuconservation})  immediately leads to the relation $T_{\nu}(a_{2}) = (a_{1}/a_{2})\,  T_{\nu} (a_{1})$.  It then follows  that 
equating~(\ref{eq:segammaR1}) and~(\ref{eq:segammaR2}) via equation~(\ref{eq:segammaconservation}) yields a $T_{\nu}/T$ temperature relation at $a=a_{2}$ that reads
\begin{equation}
\frac{T_{\nu}(a_{2})}{T(a_{2})} = \left(\frac{4}{11}\right)^{1/3}  \left( 1 + \frac{\delta s}{\left. s^{(0)} \right|_{T_{d}/m_{e} \to \infty} }\right)^{-1/3},
\end{equation}
which translates into a corrected energy ratio relation
\begin{equation}
\label{Neff_corrected_nevercoupled}
\left. \rho_\nu  \right|_{T/m_{e} \to 0}= 3 \times \frac{7}{8} \left(\frac{4}{11}\right)^{4/3} \left( 1 + \frac{\delta s}{\left. s^{(0)} \right|_{T_{d}/m_{e} \to \infty} }\right)^{-4/3} \left. \rho_\gamma  \right|_{T/m_{e} \to 0}.
\end{equation}
Then, defining $N_{\rm eff} \equiv 3+\delta N_{\rm eff}$, we find
\begin{equation}
\label{eq:neffentropycorrected}
\delta N_{\rm eff}  = 3  \left[ \left( 1 + \frac{\delta s}{\left. s^{(0)} \right|_{T_{d}/m_{e} \to \infty} }\right)^{-4/3}-1 \right] 
\end{equation}
as the corresponding change in $N_{\rm eff}$.


\subsection{Estimating $N_{\rm eff}$ from the continuity equation}\label{ODE_derivation}
\label{sec:ode}

The estimation of $N_{\rm eff}$ via entropy conservation arguments becomes poorly defined if the instantaneous decoupling approximation were relaxed as well, since a drawn-out decoupling process is generally accompanied by out-of-equilibrium energy transfer, which distorts the neutrino phase space distributions and generates entropy.  While the investigation of these effects is outside of the scope of this work, we nonetheless present the relevant equations here, as well as their modifications in the presence of non-ideal gas behaviours in section~\ref{FTQED_intro}, in preparation for a future publication.

To compute $N_{\rm eff}$ in the presence of entropy production necessitates that we solve the continuity equation for the total energy density of all particle species in the system.  Following the notation of~\cite{Mangano:2005cc,Mangano:2001iu},  the continuity equation is
\begin{align}
    \frac{\dd}{\dd{x}} \bar{\rho}_{\rm t}(x,z(x))=\frac{1}{x}\left[\bar{\rho}_{\rm t}(x,z(x))-3\bar{P}_{\rm t}(x,z(x))\right]. \label{comoving_conserv_equation}
\end{align}
Here, the dimensionless time variable is $x\equiv m_e R(t)$, where  $R(t)\equiv a(t)/[a(t_d)T_\nu (t_d)]$ is an inverse temperature parameter normalised to the nominal neutrino decoupling temperature $T_\nu(t_{d}) \equiv T_{d}$ at $t=t_{d}$;%
\footnote{This normalisation is somewhat different from that used in, e.g.,~\cite{Gariazzo:2019gyi}, where $R(t)$ had been normalised such that 
$R(t)\rightarrow1/T$ at times well before neutrino decoupling.  In the instantaneous decoupling approximation, our definition leads to $R(t)=1/T_\nu(t)$ at all times after neutrino decoupling.} 
 $\bar{\rho}_{\rm t} \equiv \rho_{\rm t} \times (x/m_e)^4$ and $\bar{P}_{\rm t} \equiv P_{\rm t} \times (x/m_e)^4$ are the dimensionless comoving total energy density and total pressure, where 
 \begin{equation}
 \begin{aligned}
 \rho_{\rm t} &\equiv \rho_{\gamma}+\rho_{e}+\rho_{\nu} \equiv \rho + \rho_{\nu}, \\
  P_{\rm t} &\equiv P_{\gamma}+P_{e}+P_{\nu}\equiv P+P_{\nu}
  \end{aligned}
  \end{equation}
 sum over all relevant quantities in the neutrino and QED sectors.   Two more associated quantities can be defined: the comoving momentum $y \equiv p R(t)$, and the rescaled photon temperature $z \equiv T(t)  R(t)$.

We are interested in the asymptotic energy density ratio $\rho_{\nu}/\rho_{\gamma}$ as $T/m_{e} \to 0$ (i.e., $x \to \infty$), 
which, under the assumptions of equilibrium within the QED sector and minimal distortions to the neutrino phase space distribution, is completely characterised by the temperature ratio $T_{\nu}/T$.  In the instantaneous decoupling limit, 
 this is exactly equivalent to solving the continuity equation~(\ref{comoving_conserv_equation}) for the asymptotic $z_{\rm fin} \equiv z(x= x_{\rm fin} \to \infty)$ using the initial conditions $z_{\rm ini} \equiv z(x=x_{\rm ini} \equiv m_{e}/T_{d}) = 1$.   It is therefore convenient to rewrite equation~(\ref{comoving_conserv_equation}) as an equation of motion for $z$;  noting that $\dd{}/\dd{x}=\partial/\partial x+(\dd{z}/\dd{x}) \partial/\partial z$, this exercise yields
\begin{align}
	\frac{\dd{z}}{\dd{x}}=&\;\frac{\frac{1}{2z^3}\Big[\frac{1}{x}(\bar{\rho}^{(0)}-3\bar{P}^{(0)})-\frac{\partial \bar{\rho}^{(0)}}{\partial x}-\frac{\dd{}}{\dd x}\bar{\rho}_\nu+\frac{1}{x}(\delta\bar{\rho}-3\delta\bar{P})-\frac{\partial \delta\bar{\rho}}{\partial x}\Big]}{\frac{1}{2z^3}\Big(\frac{\partial \bar{\rho}^{(0)}}{\partial z}+\frac{\partial \delta\bar{\rho}}{\partial z}\Big)}\label{General_ODE},
 \end{align}
where we have split the QED energy density and pressure into an ideal gas component, $\bar{\rho}^{(0)}$ and  $\bar{P}^{(0)}$, plus non-ideal gas corrections $\delta \bar{\rho}$ and  $\delta \bar{P}$.
 Evaluating the ideal gas terms explicitly yields
\begin{align}
\label{master_ODE}
	\frac{\dd{z}}{\dd{x}}=&\;\frac{\left(\frac{x}{z}\right) J (x/z)\; -\frac{\dd{}}{\dd x}\bar{\rho}_\nu
   +G_1 (x,z)}{\left(\frac{x^2}{z^2}\right) J (x/z)+Y (x/z) +\frac{2\pi^2}{15}+G_2 (x,z)},
\end{align}
with
\begin{equation}
\begin{aligned}
    \label{J}     J (\tau) & \equiv \frac{1}{\pi^2}\int_0^\infty \dd{\omega}\; \omega^2\; \frac{\exp({\sqrt{\omega^2 + \tau^2}})}{[\exp({\sqrt{\omega^2 + \tau^2}})+1]^2} , \\
           Y (\tau)& \equiv \frac{1}{\pi^2} \int_0^\infty \dd{\omega}\; \omega^4\; \frac{\exp({\sqrt{\omega^2 + \tau^2}})}{[\exp({\sqrt{\omega^2 + \tau^2}})+1]^2},
\end{aligned}
\end{equation}
and the functions
\begin{equation}
\begin{aligned}
	2z^3 G_1 (x,z) &\equiv\frac{1}{x}(\delta\bar{\rho}-3\delta\bar{P})-\frac{\partial \delta\bar{\rho}}{\partial x}, \label{G1def}	\\
				2z^3 G_2 (x,z) & \equiv \frac{\partial \delta\bar{\rho}}{\partial z}
\end{aligned}
\end{equation}
represent the yet-to-be-specified non-ideal gas corrections.

Note in equation~(\ref{master_ODE}) that the term ${\rm d} \bar{\rho}_{\nu}/{\rm d} x$ is the only total derivative remaining in the equation.  Under general circumstances, this term is proportional to the Boltzmann collision integral that dictates the weak scattering of the neutrino population. It vanishes only in the limits of (i) thermodynamic equilibrium, and (ii) negligible interaction.  The instantaneous decoupling approximation therefore corresponds to assuming that the neutrino population transits between limits  in an instant, so that ${\rm d} \bar{\rho}_{\nu}/{\rm d} x=0$ at all times.  In a realistic situation, however, we expect a nonzero ${\rm d} \bar{\rho}_{\nu}/{\rm d} x$ in the intermediate, transition regime.

Solving equation~(\ref{master_ODE}) for $G_{1}=G_{2} = 0$ (ideal gas approximation) and  ${\rm d} \bar{\rho}_{\nu}/{\rm d} x=0$ (instantaneous decoupling approximation) with the initial condition $z_{\rm ini} = 1$ set at $x_{\rm ini}= \lim_{T_{d }/m_{e}\to \infty} (m_{e}/T_{d})=0$  (ultra-relativistic approximation) yields the standard expectation
\begin{equation}
z_{\rm fin} = \left(\frac{11}{4} \right)^{1/3} \equiv \left. z^{(0)}_{\rm fin} \right|_{x_{\rm ini} = 0},
\label{eq:canonicalzfin}
\end{equation}
corresponding to $N_{\rm eff}=3$.   It then follows that the correction to $N_{\rm eff}$ induced by dropping some or all of the aforementioned assumptions is given by
\begin{equation}
\label{eq:neffzfin}
\delta N_{\rm eff} =3\left[ \Bigg( \frac{ \left. z^{(0)}_{\rm fin} \right|_{x_{\rm ini} = 0}}{z_{\rm fin}} \Bigg)^{4} -1 \right] ,
\end{equation}
where  $\delta z_{\rm fin} \equiv  z_{\rm fin}-\left. z_{\rm fin} \right|_{x_{\rm ini} = 0}$ denotes the deviation in the asymptotic $z_{\rm fin}$ value from the canonical value~(\ref{eq:canonicalzfin}).  In the limit 
${\rm d} \bar{\rho}_{\nu}/{\rm d} x=0$, equation~(\ref{eq:neffzfin}) must yield  the same result as the entropy conservation estimate~(\ref{eq:neffentropycorrected}) under the same set of assumptions.


\section{Equilibrium energy transport and the ultra-relativistic approximation}
\label{sec:nevercoupled}

As we have seen in section~\ref{sec:physical}, the canonical calculation of the neutrino-to-photon temperature ratio $T_{\nu}/T$  supposes that the electron/positron population is fully ultra-relativistic at the instant of neutrino decoupling.  Formally, this approximation requires that we push the neutrino decoupling temperature to $T_{d}/m_{e} \to \infty$ relative to the electron mass within the limited, electron/positron/photon/neutrino system considered here.
 However, because we generally expect neutrino decoupling to occur at $T_{d} \sim 1$~MeV, the condition $T_{d}/m_{e} \to \infty $ is not in fact well satisfied in reality.  Consequently, the breakdown of this approximation also constitutes the largest correction to $N_{\rm eff} = 3$, as we shall show.

\begin{figure}
\includegraphics[width=12.5cm]{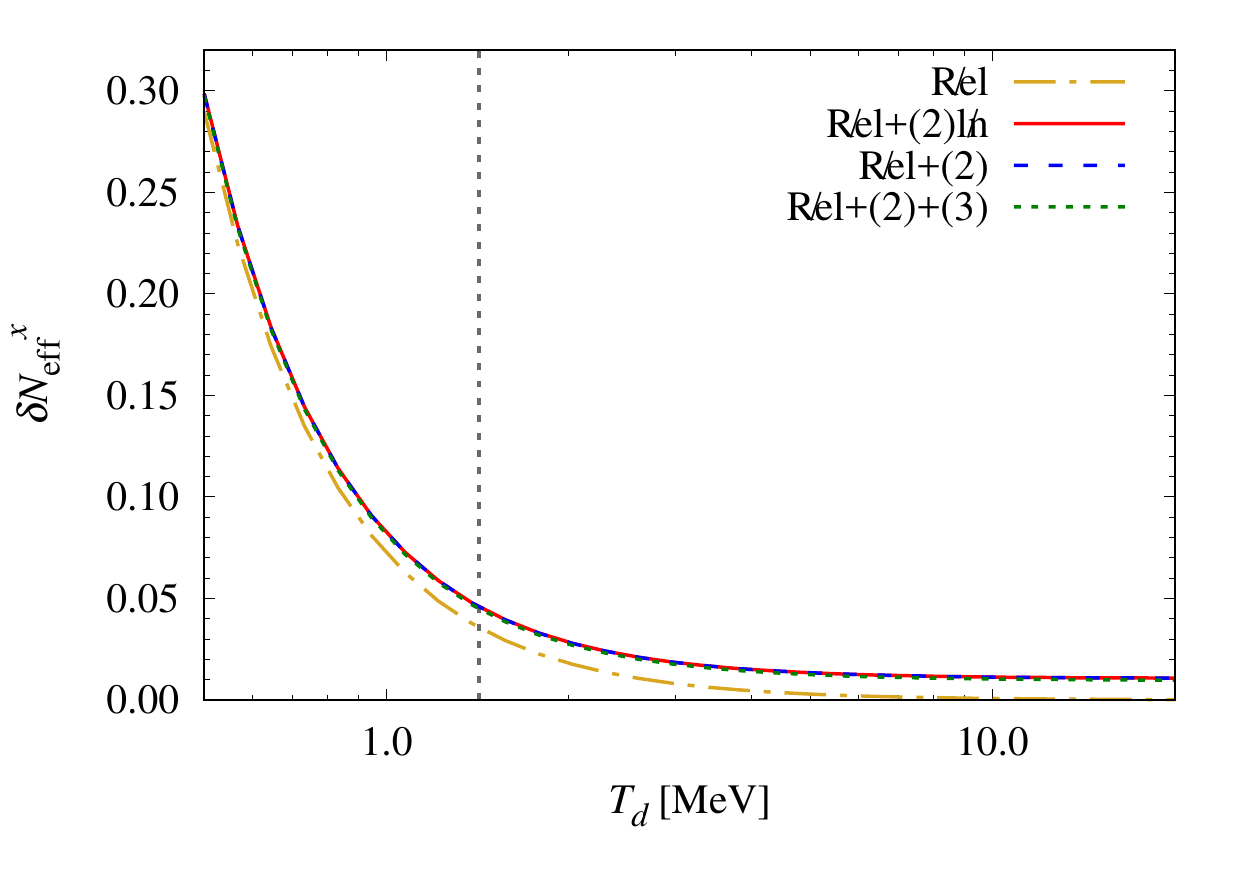}
\centering
\caption{Corrections to $N_{\rm eff}=3$ due to relaxing various assumptions, as a function of the neutrino decoupling temperature $T_d$.  The gold/dot-dash line denotes dropping the ultra-relativistic approximation ($x=\slashed{\rm Rel}$), the red/solid line includes in addition the log-independent ${\cal O}(e^2)$ FTQED correction to the QED equation of state
 ($x=\slashed{\rm Rel}+(2)\slashed{\ln}$), the blue/dashed line includes the full ${\cal O}(e^2)$ FTQED correction  ($x=\slashed{\rm Rel}+(2)$), and the green/dotted line contains FTQED corrections up to and including ${\cal O}(e^3)$  ($x=\slashed{\rm Rel}+(2)+(3)$).  The vertical grey/dotted line marks $T_{d}= 1.3453$~MeV, and we assume the decoupling between the neutrino and QED sectors to occur instantaneously.}\label{fig:Allresults}
\end{figure}

To reinstate the role of  a finite $T_{d}/m_{e}$ in the estimate of $N_{\rm eff}$, we note that dropping the ultra-relativistic approximation changes in the entropy density of the QED plasma at the time of neutrino decoupling by an amount
\begin{equation}
\label{eq:dsnnc}
\delta s^{\slashed{\rm Rel}}   =  \left. \frac{g_{e}}{2 \pi^{2} T_{d}} \int_{0}^{\infty} {\rm d} p \, p^{2} \left(E_{e} + \frac{p^{2}}{3 E_{e}} \right)  f_D (E_{e}) \right|^{T_{d}/m_{e}}_{T_{d}/m_{e} \to \infty},  \\
\end{equation}
where  $f_D(E) = [\exp(E/T)+1]^{-1}$ is the Fermi--Dirac distribution,  $E^{2}_{e} = p^{2}+m_{e}^{2}$ is the electron energy, and we remind the reader that the physical momentum $p$ scales as $p \propto a^{-1}$.   Equation~(\ref{eq:dsnnc}) can be easily evaluated numerically for any choice of neutrino decoupling temperature $T_{d}$.  Taking our estimate $T_{d} = 1.3453$~MeV from section~\ref{real_time_FTQFT} and an electron mass of $m_e = 0.511$~MeV, we find $\delta s^{\slashed{\rm Rel}}/ \left. s^{(0)} \right|_{T_{d}/m_{e} \to \infty} \simeq  -0.009859$.  It then follows simply from equation~(\ref{eq:neffentropycorrected}) that the correction to $N_{\rm eff}$ is
\begin{equation}
\delta N_{\rm eff}^{\slashed{\rm Rel}} \simeq 0.039895
\label{eq:nevercoupledcorrected}
\end{equation}
 from dropping the ultra-relativistic approximation alone.  Figure~\ref{fig:Allresults} shows $\delta N_{\rm eff}^{\slashed{\rm Rel}}$ as a function of the neutrino decoupling temperature $T_{d}$.

In terms of the continuity equation~(\ref{General_ODE}), dropping the ultra-relativistic approximation corresponds to relaxing the assumption that $x_{\rm ini} \equiv m_{e}/T_{d} = 0$ (but still keeping  $G_{1}=G_{2} =0$ and $\dd \bar{\rho}_{\nu}/\dd x=0$).
  Physically, a nonzero $\delta N_{\rm eff}$ arising from relaxing the $m_{e}/T_{d} = 0$ assumption is  simply a statement that $e^{\pm}$ annihilation is not a temporally localised event at $T \sim 0.5$~MeV.  Rather, net annihilation extends into the era before the neutrino sector even decouples from the QED plasma, which in turn enables the transfer of some of the energy originally residing with the electrons/positrons to the neutrino sector under equilibrium conditions.
Note that this phenomenon is distinct from that arising from dropping the instantaneous decoupling approximation---the two are often conflated in the literature, even though only non-instantaneous decoupling distorts the equilibrium distributions.  The magnitude of the correction~(\ref{eq:nevercoupledcorrected}) relative to established results (e.g., $N_{\rm eff} = 3.044$~\cite{Gariazzo:2019gyi}) also tells us that corrections to $N_{\rm eff}$ from genuine out-of-equilibrium energy transport effects are in fact subdominant.


\section{Non-ideal gas: Finite-temperature corrections to the QED equation of state}
\label{FTQED_intro}

At finite temperatures and densities, interacting quantum fields are known to present features not encountered at zero temperature: particles are dressed into ``quasiparticles''
and novel collective excitations appear for fermions~\cite{Klimov:1981ka,Klimov:1982bv,Weldon:1982bn,Weldon:1989ys}, gauge fields~\cite{Silin:1960pya,Fradkin1965,Kalashnikov:1979cy,Weldon:1982aq}, and scalars~\cite{Drewes:2013bfa}, i.e., plasmons, holes, etc.
The presence of the plasma does not only modify the spectrum of resonances, but also their effective widths, which can be interpreted in terms of elementary processes in the plasma that limit the quasiparticle's mean free path~\cite{Weldon:1983jn}.

In the context of $N_{\rm eff}$, while there are several entry points where finite-temperature effects can play a role, the dominant contribution is expected to come from FTQED corrections to the equation of state of the QED plasma.  These corrections have been considered in a number of previous works~(e.g., \cite{Heckler:1994tv,Lopez:1998vk,Mangano:2001iu,Grohs2016}), wherein the departure of the plasma from an ideal gas due to interactions had been invariably  formulated in terms of the acquisition of temperature-dependent masses by the interacting particles.  While at some level this view is correct, it is also easily prone to misinterpretation, particularly in the computation of bulk thermodynamics quantities, as we shall show in section~\ref{sec:mistakes}.

A more fool-proof calculation should begin with the grand canonical partition function~$Z$ of the QED plasma at finite temperatures, for which  
a systematic expansion of the Helmholtz free energy $F \equiv -T \ln Z$ in powers of the QED coupling constant~$e$ (i.e., the elementary electric charge),
\begin{equation}\label{partition_function_expansion}
    \ln Z = \ln Z^{(0)} + \ln Z^{(2)}+ \ln Z^{(3)} + \; \cdots
\end{equation}
where $\ln Z^{(n)} \propto e^{n}$, is known to $n=3$ for arbitrary $m_{e}$ and $\mu$~\cite{Kapusta:2006pm} and to $n=5$ in the $m_{e}=\mu=0$~limit~\cite{Coriano,Parwani:1994xi}.%
\footnote{We note in passing that FTQED corrections do not always appear as integer powers of $e$, because resummation effects can generate logarithmic corrections as well (see, e.g., \cite{Kapusta:2006pm}).}
Then, at each order~$n$, standard thermodynamics relations can be used to deduce from~$\ln Z^{(n)}$ the corresponding  pressure~$P^{(n)}$, energy density $\rho^{(n)}$, and entropy density~$s^{(n)}$:
\begin{align}
P^{(n)} &= \frac{T}{V} \ln Z^{(n)}, \label{Pressure} \\
\label{from_P_to_rho}
\rho^{(n)} & = \frac{T^{2}}{V} \frac{\partial \ln Z^{(n)}}{\partial T} = -P^{(n)} + T \frac{\partial P^{(n)}}{\partial T}, \\
s^{(n)} &= \frac{1}{V}
\frac{\partial \left[T \ln Z^{(n)}\right]}{\partial T}= \frac{\rho^{(n)} + P^{(n)}}{T}, 
\end{align}
where $T$ and $V$ are the temperature and volume of the system respectively.  At zeroth order the pressure and energy density are
\begin{align}
\label{P0}
  P^{(0)} & = \frac{T}{\pi^2}\int_0^\infty \dd{p}\;p^2\;\ln\bigg[\frac{(1+\mathrm{e}^{-E_e/T})^2}{(1-\mathrm{e}^{-E_\gamma/T})}\bigg], \\
\label{rho0}
    \rho^{(0)} & = \frac{1}{ \pi^2 }\int_0^\infty \dd{p} \; p^2 \left[\frac{2 E_e}{e^{E_e/T}+1} + \frac{E_\gamma}{e^{E_\gamma/T}-1} \right],
\end{align}
which are simply what we should expect of an ideal gas of photons and electrons/positrons, with $E_{\gamma}=p$ and $E_{e}^{2}=p^{2}+m_e^{2}$.


\subsection{$\mathcal{O}(e^2)$ FTQED}

\begin{figure}[t]
    \centering
\includegraphics[width=5.cm]{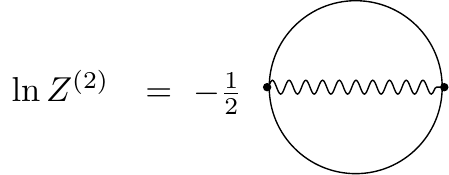}
 \caption{Diagrammatic expression for the ${\cal O}(e^2)$ correction to the FTQED partition function.}
 \label{diagram_lnZ2}
\end{figure}

Corrections at $\mathcal{O}(e^2)$ to the QED partition function are represented diagrammatically by the  two-loop diagram shown in figure~\ref{diagram_lnZ2}.  This diagram has been evaluated by many, and the general result for $\ln Z^{{(2)}}$ can be found in, e.g., equation~(5.58) of~\cite{Kapusta:2006pm}.  For an isotropic and $CP$-symmetric medium such as that under consideration, the expression simplifies to
\begin{equation}
\begin{aligned}
P^{(2)}= \frac{T}{V}\ln Z^{(2)} = & - \frac{e^2 T^2}{12\pi^2}\int_0^{\infty} \dd{p} \; \frac{ p^2}{E_p}n_{D}-\frac{e^2}{8\pi^4}\left(\int_0^{\infty} \dd{p}\; \frac{ p^2}{E_p}n_{D}\right)^2 \\
& \qquad + \frac{e^2 m_{e}^{2}}{16 \pi^4 } \iint_0^\infty \dd{p} \;\dd{\tilde{p}} \;\frac{p \tilde{p}}{E_p E_{\tilde{p}}}\;\ln\left|\frac{p+\tilde{p}}{p-\tilde{p}}\right| \;  n_D\tilde{n}_D,
\label{e^2_pressure}
\end{aligned}
\end{equation}
where we have  defined $E_{p}^{2}=p^{2} + m_{e}^{2}$, $n_{D} \equiv 2 f_D (E_{p}) = 2 [\exp(E_{p}/T)+1]^{-1}$, and $\tilde{n}_{D} \equiv 2 f_D (E_{\tilde{p}})$.   In the limit $T/m_{e}\to \infty$, equation~(\ref{e^2_pressure}) evaluates to  $P^{(2)} = -5 e^{2} T^{4}/288$.

As we shall discuss in detail in section~\ref{sec:mistakes},  with the exception of~\cite{Grohs2016}, FTQED corrections applied to precision $N_{\rm eff}$ calculations to date are essentially equivalent to using only the first two, ``log-independent'' terms of equation~(\ref{e^2_pressure}), following the recipe laid down in~\cite{Mangano:2001iu}, itself based on~\cite{Heckler:1994tv}.  The third, ``log-dependent'' term is usually deemed too small to warrant detailed investigation on its impact on $N_{\rm eff}$~\cite{Lopez:1998vk,Mangano:2001iu}.
  We  examine both contributions in the following.


\subsubsection{Log-independent contribution ($\slashed{\ln}$)}
\label{sec:2login}

Using only the first two terms of equation~\eqref{e^2_pressure}, we find the energy and entropy density corrections
\begin{align}
  {\rho}^{(2)\slashed{\ln}} = & -\frac{e^2 T^2 }{12 \pi^2}\int_0 ^\infty \dd{p} \frac{p^2}{E_p} \left(n_D + T \partial_T n_D\right) + \frac{e^2}{8\pi^4}\left(\int_0^{\infty} \dd{p} \frac{ p^2}{E_p}n_D\right)^2 \nonumber \\    & \qquad  \quad -\frac{e^2}{4\pi^4} \left(\int_0^\infty \dd{p} \frac{p^2}{E_p} \; n_D\right)
\left( \int_0^\infty     \dd{p}  \frac{p^2}{E_p} \; T \partial_T n_D \right), \\
  s^{(2)\slashed{\ln}}  = & -\frac{e^2 T }{12 \pi^2}\int_0 ^\infty \dd{p} \frac{p^2}{E_p} \left(2 n_D + T \partial_T n_D\right) -\frac{e^2}{4\pi^4} \left(\int_0^\infty \dd{p} \frac{p^2}{E_p} \; n_D\right)
\left( \int_0^\infty     \dd{p}  \frac{p^2}{E_p} \;  \partial_T n_D \right),
\end{align}
which, in the limit $T/m_{e} \to \infty$, give ${\rho}^{(2)\slashed{\ln}}= - 5 e^{2} T^{4}/96$ and $s^{(2)\slashed{\ln}} = - 5 e^{2} T^{3}/72$, respectively.  In the general case, however, the phase space integrals need to be evaluated numerically.

\begin{figure}
\includegraphics[width=12.5cm]{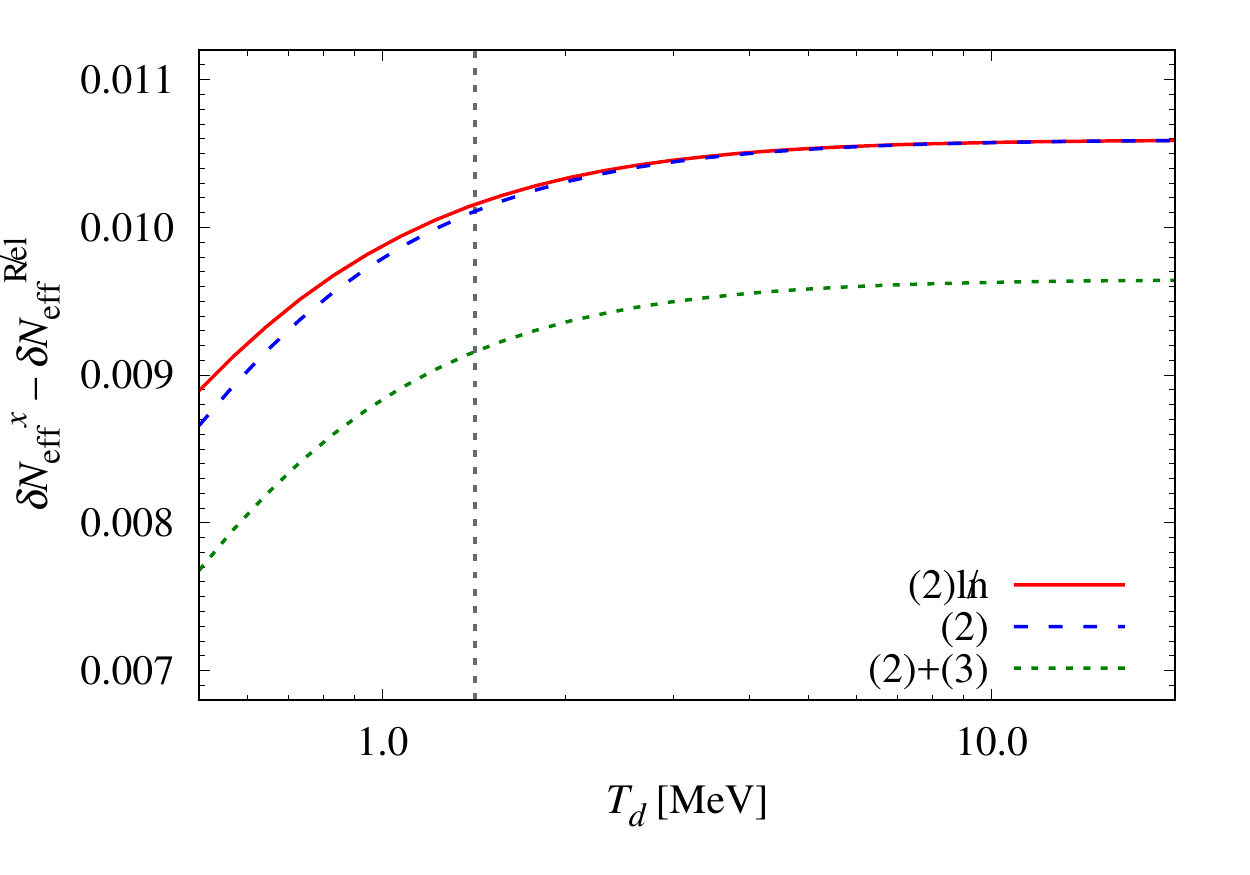}
\centering
\caption{Effective corrections to $N_{\rm eff}=3$ due to FTQED effects on the QED equation of state alone (i.e., with the $\slashed{\rm Rel}$ contribution subtracted), 
as a function of the neutrino decoupling temperature $T_d$.   The red/solid line denotes the log-independent ${\cal O}(e^2)$ contribution
 ($x=\slashed{\rm Rel}+(2)\slashed{\ln}$), the blue/dashed line the full ${\cal O}(e^2)$ correction  ($x=\slashed{\rm Rel}+(2)$), and the green/dotted line includes corrections up  to and including ${\cal O}(e^3)$  ($x=\slashed{\rm Rel}+(2)+(3)$).  The vertical grey/dotted line marks $T_{d}=1.3453$~MeV, and we assume the decoupling between the neutrino and QED sectors to occur instantaneously.}\label{fig:Allresultsthermal}
\end{figure}

To compute the corresponding change in $N_{\rm eff}$  via entropy conservation arguments, we first identify $\delta s$ of equation~(\ref{eq:neffentropycorrected}) with
\begin{equation}
\delta s = \delta s^{\slashed{\rm Rel}} + \left.s^{(2)\slashed{\ln}} \right|_{T=T_d},
\end{equation}
where $\delta s^{\slashed{\rm Rel}}$ is the change in the QED entropy density at $T=T_{d}$ from dropping the ultra-relativistic approximation given in equation~(\ref{eq:dsnnc}).  Evaluating $\delta s$ at $T_{d} = 1.3453$~MeV, we find  $\delta s/ \left. s^{(0)} \right|_{T_{d}/m_{e} \to \infty} \simeq  -0.012324$, and hence a correction of
\begin{equation}
\label{eq:nnc2}
\delta N_{\rm eff}^{\slashed{\rm Rel}+(2)\slashed{\ln}} \simeq 0.050015,
\end{equation}
where we have used a fine structure constant value $\alpha \equiv e^2/4 \pi = 1/137$.  
Subtracting from~(\ref{eq:nnc2}) the correction (\ref{eq:nevercoupledcorrected}) due to dropping the ultra-relativistic approximation, we find a net ${\cal O}(e^{2})$ log-independent FTQED correction of
\begin{equation}
\delta N_{\rm eff}^{(2)\slashed{\ln}} = \delta N_{\rm eff}^{\slashed{\rm Rel}+(2)\slashed{\ln}}-  \delta N_{\rm eff}^{\slashed{\rm Rel}}  \simeq 
0.010121
\label{eq:e2nolog}
\end{equation}
for  $T_{d} = 1.3453$~MeV, which should be compared  with the oft-quoted  $\delta N_{\rm eff}^{(2)\slashed{\ln}}  \simeq 0.010594$~\cite{Lopez:1998vk} computed from the 
same FTQED correction but in the limit $T_{d}/m_{e} \to \infty$. 
Figure~\ref{fig:Allresults} shows $\delta N_{\rm eff}^{\slashed{\rm Rel}+(2)\slashed{\ln}}$ for a range of neutrino decoupling temperatures $T_{d}$; figure~\ref{fig:Allresultsthermal} shows the correction $\delta N_{\rm eff}^{(2)\slashed{\ln}}$ alone (i.e., with the $\slashed{\rm Rel}$ contribution subtracted from  $\delta N_{\rm eff}^{\slashed{\rm Rel}+(2)\slashed{\ln}}$).

Identifying $\delta \bar{\rho} =  {\rho}^{(2)\slashed{\ln}}  \times (x/m_e)^4$ and  $\delta \bar{P} =  {P}^{(2)\slashed{\ln}} \times (x/m_e)^4$, equation~\eqref{G1def} allows us to evaluate the  corresponding $G_1(x/z)$ and $G_2(x/z)$  functions for the continuity equation~(\ref{General_ODE}):
\begin{align}
  \label{G1Mang}  G^{(2)\slashed{\ln}}_{1} (\tau)  =& \frac{e^2}{2}\bigg[\frac{1}{\tau}\bigg(\frac{K(\tau)}{3}+2K(\tau)^2-\frac{J(\tau)}{6}-J(\tau)K(\tau)\bigg) \nonumber \\
    & \qquad +\frac{K'(\tau)}{6}-K(\tau)K'(\tau)+\frac{J'(\tau)}{6}+J'(\tau)K(\tau)+J(\tau)K'(\tau)\bigg], \\
       G^{(2)\slashed{\ln}}_{2}(\tau) = & \frac{e^2}{2}\bigg[2 J(\tau) K(\tau)  + \left(\tau J'(\tau) + \frac{1}{\tau}Y'(\tau)\right)\left(K(\tau)+\frac{1}{6}\right)  - J(\tau)^2 -\frac{(K(\tau)+J(\tau))}{3}
     \bigg], \label{G2_Mangano} 
  \end{align}
where $(\cdots )' \equiv \partial_\tau (\cdots)$, the new function $K(\tau)$ is defined as
\begin{equation}
\label{eq:K}
   K (\tau)\equiv \frac{1}{\pi^2}\int_0^\infty \dd{\omega}\; \frac{\omega^2}{\sqrt{\omega^2+\tau^2}}\; \frac{1}{\exp{(\sqrt{\omega^2+\tau^2})}+ 1},
\end{equation}
while  $J(\tau)$ and $Y(\tau)$ are given in equation~(\ref{J}).

Observe that the expression~(\ref{G1Mang})  is identical to equation~(18) of~\cite{Mangano:2001iu}.  Our expression for $G^{(2)\slashed{\ln}}_2$,  on the other hand, is formally different from the result reported in equation~(19) of~\cite{Mangano:2001iu}.  However, we have checked that, numerically, equation~(\ref{G2_Mangano}) and the corresponding expression in~\cite{Mangano:2001iu}  are identical to within machine precision for a range of $\tau$ values.  We therefore do not dwell further on reproducing the formal result of~\cite{Mangano:2001iu}  exactly, but only note that, with the exception of~\cite{Grohs2016}, all recent precision calculations of $N_{\rm eff}$~\cite{Mangano:2005cc,deSalas,Birrell:2014uka,Gariazzo:2019gyi} account for FTQED effects on the QED equation of state by solving the continuity equation~(\ref{General_ODE}) modified with the corrections~(\ref{G1Mang}) and~(\ref{G2_Mangano}) (or their equivalent presented in~\cite{Mangano:2001iu}).%
\footnote{The oft-quoted $\delta N_{\rm eff}^{(2)\slashed{\ln}}  \simeq 0.010594$ corresponds to $z_{\rm fin} \simeq 1.39979$ via equation~(\ref{eq:neffzfin}), and can be obtained from a numerical solution of the continuity equation~(\ref{General_ODE}) with the initial conditions set at $x_{\rm ini} = 0$ as per definition.  The number quoted in~\cite{Mangano:2001iu,deSalas}, $\zfin=1.39975$, corresponds to setting the initial conditions at $x_{\rm ini} =0.0341 $, or, equivalently, $T_{d}=15$~MeV.}
We defer the discussion of the ``alternative'' FTQED implementation of~\cite{Grohs2016} to section~\ref{sec:mistakes}.


\subsubsection{Logarithmic contribution ($\ln$)}
\label{sec:o2log}
 
The energy and entropy density corrections corresponding to the third, logarithmic term of equation~\eqref{e^2_pressure} are, respectively,
\begin{align}
    \rho^{(2)\ln} & = \frac{e^2 m_{e}^2 }{16 \pi^4}\iint_0^\infty \dd{p} \; \dd{\tilde{p}} \; \frac{p \tilde{p}}{E_p E_{\tilde{p}}} \ln\left|\frac{p+\tilde{p}}{p-\tilde{p}}\right| \; n_{D} \left(2T \partial_T \tilde{n}_D - \tilde{n}_D\right), \\
     s^{(2),\ln} & = \frac{e^2 m_{e}^2}{8 \pi^4} \iint_0^\infty \dd{p} \; \dd{\tilde{p}}\;  \frac{p \tilde{p}}{E_p E_{\tilde{p}}} \ln\left|\frac{p+\tilde{p}}{p-\tilde{p}}\right|  \; n_{D} \partial_T \tilde{n}_D, 
\end{align}
both of which are vanishing in the $T/m_{e} \to \infty$ limit.  It is usually argued that even at a finite $T/m_{e}$, the magnitude of this logarithmic contribution is less than 10\% of the log-independent term of the previous section~\cite{Mangano:2001iu}, and as such is negligible.

To assess this claim and to estimate the corresponding correction to $N_{\rm eff}$, we identify the entropy deviation at neutrino decoupling with
\begin{equation}
\delta s = \delta s^{\slashed{\rm Rel}} + \left.s^{(2)\slashed{\ln}} \right|_{T=T_d} + \left.s^{(2)\ln} \right|_{T=T_d}.
\end{equation}
Then, for a decoupling temperature $T_{d} = 1.3453$~MeV, we find  $\delta s/ \left. s^{(0)} \right|_{T_{d}/m_{e} \to \infty} \simeq -0.012312$, leading to a correction to $N_{\rm eff}$ of
\begin{equation}
\delta N_{\rm eff}^{\slashed{\rm Rel}+(2)} \simeq 0.049965,
\end{equation}
or, equivalently, a net ${\cal O}(e^{2})$ FTQED contribution of 
\begin{equation}
\delta N_{\rm eff}^{(2)} = \delta N_{\rm eff}^{\slashed{\rm Rel}+(2)}-  \delta N_{\rm eff}^{\slashed{\rm Rel}}  \simeq 0.010070,
\end{equation}
of which
\begin{equation}
\delta N_{\rm eff}^{(2)\ln} = \delta N_{\rm eff}^{(2)} - \delta N_{\rm eff}^{(2)\slashed{\ln}} \simeq -0.000050,
\end{equation}
or about 0.5\% of $\delta N_{\rm eff}^{(2)}$, comes from the ${\cal O}(e^{2})$ logarithmic term alone.  Thus, the logarithmic contribution appears to be even less significant than previously envisaged.

Nonetheless, for completeness we report here the associated $G_{1}$ and $G_{2}$  functions in the continuity equation:
\begin{align}
    G_{1}^{(2)\ln}(\tau)=&\,\frac{e^2 x}{16 \pi^4 z^3}\iint_0^\infty \dd{y}\; \dd{\tilde{y}} \frac{y}{\sqrt{y^{2}+x^{2}}} \frac{\tilde{y}}{\sqrt{\tilde{y}^{2}+x^{2}}}
    \ln\left|\frac{y + \tilde{y}}{y-\tilde{y}}\right| \bigg\{ -n_D\tilde{n}_D -z n_D\partial_z \tilde{n}_D\nonumber \\
   &\qquad \qquad  \qquad \qquad  -x \left[ z\left(\partial_x n_D\partial_z\tilde{n}_D + n_D\partial_x\partial_z \tilde{n}_D\right)- n_D\partial_x  \tilde{n}_D \right] \\
 &   \qquad \qquad \qquad \qquad \qquad +\frac{x^2(y^{2}+x^{2} + \tilde{y}^{2}+x^{2})}{2 (y^{2}+x^{2})( \tilde{y}^{2}+x^{2})} \left(2z n_D \partial_z \tilde{n}_D- n_D \tilde{n}_D\right)\bigg\},\nonumber \\
   G_{2}^{(2)\ln}(\tau)=& \,\frac{e^2 x^2}{16 \pi^4 z^2}\iint_0^{\infty}\dd{y} \; \dd{\tilde{y}}
   \frac{y}{\sqrt{y^{2}+x^{2}}} \frac{\tilde{y}}{\sqrt{\tilde{y}^{2}+x^{2}}}
\ln\left|\frac{y + \tilde{y}}{y-\tilde{y}}\right|\partial_z\left( n_D\partial_z \tilde{n}_D\right).
\end{align}
Solution of the continuity equation including these corrections for a range of neutrino decoupling temperatures, shown in figures~\ref{fig:Allresults} and~\ref{fig:Allresultsthermal}, 
again confirms the insignificance of the ${\cal O}(e^{2})$ logarithmic term relative to both the total ${\cal O}(e^{2})$  FTQED contribution, as well as to our four-significant-digit accuracy goal.    We therefore conclude that the ${\cal O}(e^{2})$ logarithmic contribution can indeed be considered optional.


\subsection{$\mathcal{O}(e^3)$ FTQED}
\label{sec:o3}

\begin{figure}[t]
    \centering
\includegraphics[width=12cm]{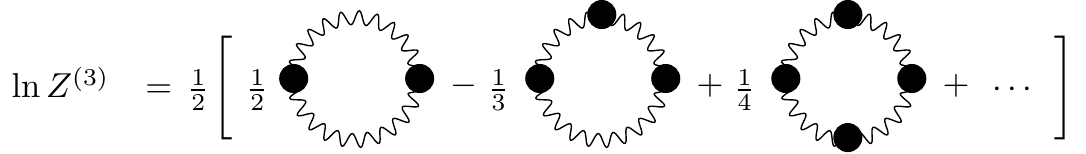}
 \caption{Diagrammatic expression for the ${\cal O}(e^3)$ correction to the FTQED partition function.  The infinite series of ring/plasmon diagrams, in which the filled black circles denote the photon self-energy at finite temperature, represents resummation to all orders.}
 \label{ring_diagrams}
\end{figure}

Unlike what standard perturbation theory would lead us to expect, the next correction to the partition function $\ln Z$  is not $\mathcal{O}(\alpha^2) = \mathcal{O}(e^4)$, but rather  $\mathcal{O}(e^3)$,
 which stems from the resummation of ring diagrams to all orders as shown in
 figure~\ref{ring_diagrams}.    
The resummation of self-energy insertions in the photon lines in figure \ref{ring_diagrams} shifts the pole of the propagator in a temperature-dependent manner. This shift gives the photon a temperature-dependent effective mass that can be understood as the result of screening in the plasma. 
This effective mass regularises the infrared divergence of the Bose--Einstein distribution for massless particles, bringing a power of $e$ into the denominator loop integrals, which explains the odd power of $e$.

The pressure correction at this order reads~\cite{Kapusta:2006pm} 
\begin{equation}\label{e^3_pressure}
    P^{(3)}= \frac{T}{V}\ln Z^{(3)}= \frac{e^3 T}{12 \pi^4} I^{{3/2}}(T),
\end{equation}
where
\begin{equation}
 I(T)=\int_0^\infty \; \dd{p}\left( \frac{p^2 + E_p^2}{E_p}\right) n_D,
\end{equation}
and we note in passing that in the nonrelativistic limit and using Maxwell--Boltzmann statistics, $P^{(3)}$ is identically the Debye--H\"uckel pressure correction due to the screening of the static Coulomb potential~\cite{Heckler:1994tv,Kapusta:2006pm}.
The corresponding energy and entropy density corrections are
\begin{align}
\rho^{(3)} & =\frac{e^3 T^2 }{8 \pi^4}I^{1/2}\partial_T I, \\
s^{(3)} & = \frac{e^3}{24 \pi^4} \left(2 I^{{3/2}} + 3 I^{1/2} T \partial_T I \right).
\end{align}
In the limit $T/m_{e} \to \infty$,  these expressions yield $P^{(3)} = e^{3}T^{4}/(36 \sqrt{3} \pi)$, $\rho^{(3)} = e^{3}T^{4}/(12 \sqrt{3} \pi)$, and $s^{(3)} = e^{3}T^{3}/(9 \sqrt{3} \pi)$.

Following the same procedure as before and writing
\begin{equation}
\delta s = \delta s^{\slashed{\rm Rel}} + \left.s^{(2)\slashed{\ln}} \right|_{T=T_d} + \left.s^{(2)\ln} \right|_{T=T_d} + \left. s^{(3)} \right|_{T=T_d},
\end{equation}
we find immediately $\delta s/ \left. s^{(0)} \right|_{T_{d}/m_{e} \to \infty} \simeq -0.012081$ for $T_{d}= 1.3453$~MeV.  Then, equation~(\ref{eq:neffentropycorrected}) yields a correction 
\begin{equation}
\delta N_{\rm eff}^{\slashed{\rm Rel}+(2)+(3)} \simeq 0.049013,
\end{equation}
or, equivalently, a net ${\cal O}(e^{2})+{\cal O}(e^{3})$ FTQED contribution of
\begin{equation}
\delta N_{\rm eff}^{(2)+(3)} = \delta N_{\rm eff}^{\slashed{\rm Rel}+(2)+(3)}-  \delta N_{\rm eff}^{\slashed{\rm Rel}}  \simeq 0.009119,
\end{equation}
of which 
\begin{equation}
\delta N_{\rm eff}^{(3)}= \delta N_{\rm eff}^{\slashed{\rm Rel}+(2)+(3)}-\delta N_{\rm eff}^{\slashed{\rm Rel}+(2)} \simeq -0.000952
\label{eq:dNe3ftqed}
\end{equation}
stems from the ${\cal O}(e^{3})$ correction of this section.  The  associated $G_{1}$ and $G_{2}$ functions  are
\begin{equation}
\begin{aligned}
     G_1^{(3)}(\tau)=&\,\frac{e^3}{4\pi}\bigg(K+\frac{\tau^2}{2}k\bigg)^{1/2}  \\
 & \times   \Bigg[\frac{1}{\tau}\Big(2J-4K\Big)-2J'-\tau^2j' -\tau\Big( 2k+j\Big)   
-\frac{\big(2J+\tau^2j\big)\big(\tau\big(k-j\big)+K'\big)}{2\big(2K+\tau^2k\big)}\Bigg], \label{G23} \\
     G_2^{(3)}(\tau)=&\,\frac{e^3}{4\pi}\Bigg(K+\frac{\tau^2}{2}k\Bigg)^{1/2}\Bigg[\frac{\big(2J+\tau^2j\big)^2}{2\big(2K+\tau^2k\big)}-\frac{2}{\tau} Y'      -\tau\Big(3J'+\tau^2j'\Big)\Bigg],
\end{aligned}
\end{equation}
with special functions
\begin{equation}
\begin{aligned}
  k (\tau)\equiv&\;\frac{1}{\pi^2}\int_0^\infty \dd{\omega}\; \frac{1}{\sqrt{\omega^2+\tau^2}}\; \frac{1}{\exp{(\sqrt{\omega^2+\tau^2})}+ 1}, \\
  \label{j}      j (\tau)\equiv &\;\frac{1}{\pi^2}\int_0^\infty \dd{\omega}\;  \frac{\exp({\sqrt{\omega^2 + \tau^2}})}{(\exp({\sqrt{\omega^2 + \tau^2}})+1)^2},
\end{aligned}
\end{equation}
and  $J(\tau)$, $Y(\tau)$, and $K(\tau)$ given in equations~(\ref{J}) and~(\ref{eq:K}).
Figures~\ref{fig:Allresults} and~\ref{fig:Allresultsthermal} show the correction to $N_{\rm eff}$ as a function of the neutrino decoupling temperature $T_{d}$.

There are two immediately notable points about the result~(\ref{eq:dNe3ftqed}) and figure~\ref{fig:Allresultsthermal}.  Firstly, the correction to $N_{\rm eff}$ from ${\cal O}(e^{3})$ FTQED is negative, i.e., it has the effect of cancelling to a small extent the ${\cal O}(e^{2})$ FTQED correction discussed in section~\ref{sec:2login} that has been used in all precision $N_{\rm eff}$ calculations thus far. While it is well known that the ${\cal O}(e^{3})$  term in the equation of state is related to Debye screening, it is not obvious on the microphysics level why this term should shift $N_{\rm eff}$ in the direction it does.
 However, the dependence on $e$ clearly suggests that it comes from contributions to the pressure and energy density from infrared-sensitive loop corrections that have been regularised by the effective photon mass.

Secondly, the  ${\cal O}(e^{3})$ FTQED  correction is, across  a broad  range of plausible $T_{d}$, not only sizeable relative to our four-significant-digit accuracy goal, but importantly, also larger than---or, at least, comparable to---the change induced in $N_{\rm eff}$ between including and excluding neutrino  oscillations in the full neutrino energy transport calculation: reference~\cite{Mangano:2005cc} reports, for example, a deviation of up to $\delta N_{\rm eff}^{\rm osc}\simeq -0.0004$ when oscillations are included, while~\cite{deSalas} finds that oscillations incur a shift in the fifth significant digit.
Neutrino  oscillations have long been considered standard ingredient in the canon of precision $N_{\rm eff}$ calculations~\cite{deSalas,Gariazzo:2019gyi};  in light of its relative importance, it is  only consistent that  the said ${\cal O}(e^{3})$ FTQED correction also be incorporated in future calculations as a standard input.

\subsection{$\mathcal{O}(e^4)$ FTQED}

For completeness and to provide a measure of the theoretical uncertainty,  we estimate also the $\mathcal{O}(e^4)$ FTQED correction in the $T/m_{e} \to \infty$ limit from the corresponding pressure correction~\cite{Coriano}:
\begin{align}
    \left. P^{(4)} \right|_{T/m_e \to \infty} =     \frac{T}{V} \left. \ln Z^{(4)} \right|_{T/m_e \to \infty}\simeq -0.0611\frac{e^4}{\pi^6} T^4, 
\end{align}
where, for simplicity, we have ignored a $5\%$ uncertainty in the numerical prefactor and set the renormalisation scale to be equal to the temperature.   In the same limit, thermodynamics relations yield $\rho^{(4)} \simeq -0.1833 \, e^{4} T^{4}/\pi^{6}$ and $s^{(4)} \simeq -0.2444 \, e^{4}T^{3}/\pi^{6}$, from which we can immediately estimate via entropy conservation~(\ref{eq:neffentropycorrected}) a corresponding correction of
\begin{equation}
\delta N_{\rm eff}^{(4)} \simeq 3.5 \times 10^{-6}.
\end{equation}
Barring the realisation of Hubble-volume surveys, which, under idealised situations, may have the capacity to probe  $N_{\rm eff}$ at the $10^{-6}$ level~\cite{Diacoumis:2018nbq}, cosmological surveys of the near-future are unlikely to reach a level of sensitivity for which corrections of this magnitude would play a role.  We therefore conclude that the $\mathcal{O}(e^4)$ FTQED correction to the QED equation of state is, for the time-being, unnecessary.


\subsection{Avoiding mistakes}
\label{sec:mistakes}

One of the motivations behind this work was the recent result of~\cite{Grohs2016}, whose reported value of $\Neff=3.052$ has apparently as much as  $\delta N_{\rm eff} \simeq 0.02$  attributed to FTQED effects on the QED equation of state.  Clearly, this result is twice as large as our findings using well-established FTQED corrections to the QED partition function and standard thermodynamic relations.  It is also similarly discrepant with previous calculations by others that followed a more heuristic description in terms of thermal mass corrections.  As this is a large change from the consensus, it warrants some detailed exploration.


\subsubsection{Thermal mass interpretation of the $\mathcal{O}(e^2)$ FTQED effects}
\label{thermal_mass_interpretation}

In order to pinpoint what might have gone wrong in the calculation of~\cite{Grohs2016}, let us first review the ``thermal mass'' narrative, due to~\cite{Heckler:1994tv} largely as a {\it side remark},  that has become commonplace and, unfortunately, over- or even misinterpreted as a {\it definition} of finite-temperature effects
in the past two decades of discussions of FTQED corrections to $N_{\rm eff}$.

Consider the expression~\eqref{P0} for the zeroth-order (i.e., ideal gas) pressure~$P^{(0)}$.  To compute finite-temperature corrections, section~V of reference~\cite{Heckler:1994tv} instructs us to replace in $P^{(0)}$ the vacuum photon and electron/positron dispersion relations with the temperature-dependent in-medium ones, i.e.,  
\begin{equation}
\begin{aligned}
  E_{\gamma}^{2}(p) &\to E_{\gamma}^{2}(p,T)= p^{2}  + \delta  m_{\gamma}^{2}(T), \\
  E_{e}^{2} (p) &\to E_{e}^{2}(p,T)= p^{2} + m_{e}^{2} + \delta  m_{e}^{2}(p,T),
 \end{aligned}
 \end{equation} 
where~\cite{Weldon:1982aq,Donoghue:1983qx,Fornengo:1997wa}
\begin{align}
       \delta m^2_\gamma(T) 
       &= \frac{e^{2}}{\pi^{2}} \int_0^\infty \dd{p}\;\frac{p^2}{E_{p}} n_{D}\label{comoving_photon_Tmass}, \\
        \delta m_e^2(p,T)        & = \frac{e^{2} T^2}{6} + \frac{e^{2}}{2 \pi^{2}}\int_0^\infty \dd{\tilde{p}}\;  \frac{\tilde{p}^2}{E_{\tilde{p}}} \; \tilde{n}_{D} 
  -\frac{m_e^2 e^{2}}{4 \pi^{2}p}\int_0^\infty \dd{\tilde{p}}\;\frac{\Tilde{p}}{E_{\tilde{p}}} \ln{\left|\frac{p+\tilde{p}}{p-\tilde{p}} \right|} \;\tilde{n}_{D} \label{eq:meth}
      \end{align}
are the $\mathcal{O}(e^2)$ photon and electron/positron thermal mass corrections, respectively, acquired in a QED plasma at finite temperature. Note that these ``masses'' are momentum-dependent, which simply reflects the fact that the dispersion relations for quasiparticles in a medium are in general complicated functions of the momentum.
We refer to this step as the ``quasiparticle picture'' assumption, and denote the pressure thus obtained $P^{\rm qp}$.

 The next step is to Taylor-expand $P^{\rm qp}$ in powers of $e$, which, for the thermal mass corrections~(\ref{comoving_photon_Tmass}) and (\ref{eq:meth}), yields a series
\begin{equation}
\label{eq:pqp}
    P^{\rm qp} = P^{(0)} +  P^{\rm qp}_{e^{2}} + P^{\rm qp}_{e^{4}} + \; \cdots,
\end{equation}
where $P^{\rm q}_{e^{2}}$ is the ${\cal O}(e^{2})$ term of interest. Multiplying $ P^{\rm qp}_{e^{2}}$ by  $1/2$, it is straightforward to show that $(1/2) P^{\rm qp}_{e^{2}}= P^{(2)}$, where  $P^{(2)}$ is identically the $\mathcal{O}(e^{2})$ pressure correction~(\ref{e^2_pressure}).%
\footnote{We emphasise that while the Taylor expansion is a necessary step in the procedure of~\cite{Heckler:1994tv} in order to recover $P^{(2)}$, it is in general not necessary for the self-consistency of the quasiparticle picture {\it per se}.}

While this procedure works at $\mathcal{O}(e^{2})$ and does provide some level of physical insight into how thermodynamic quantities depart from their ideal gas limits in the presence of interactions,  this quasiparticle picture is however incomplete and certainly highly prone to misinterpretation as a recipe for the computation of higher-order FTQED corrections to bulk thermodynamics quantities.

Firstly, the procedure requires an \textit{ad hoc} insertion of a factor $1/2$ in order to produce a $\mathcal{O}(e^{2})$ pressure correction that matches $P^{(2)}$. In the field theoretical description this factor has a well-posed diagrammatic origin---it corresponds to symmetry factors arising from the number of different ways in which the electron lines can be cut to reduce the two-loop $\ln Z^{(2)}$ to the one-loop self-energy that gives the quasiparticles their thermal masses and widths~\cite{Heckler:1994tv}. 
Classically, this factor can be understood by noticing that the deviation from the ideal gas behaviour comes from the interactions between the particles in the plasma:
The screening that a single particle experiences from the interaction with its neighbours can effectively be described in the quasiparticle picture.  However, assigning a thermal mass to every particle in the plasma double-counts the interaction energy between each pair of particles.
This classical interpretation immediately suggests that the factor should take values different from $1/2$ for higher-order terms. 
Indeed, as emphasised in~\cite{Heckler:1994tv}, the prefactor needs to fixed order by order,
which further detracts from the procedure's computational advantage, especially in light of the ready availability of the FTQED partition function.

 Secondly, the procedure described above entirely ignores the fact that there are collective excitations in the plasma that can behave like well-defined quasiparticles, such as plasmons, transversal photons, or fermionic holes. 
 Collective excitations are infrared phenomena that rely on coherent behaviours across inter-particle distances; as such they are typically only relevant for modes with soft momenta with $p \ll T$, and they do not affect the $\mathcal{O}(e^2)$ results. However, collective excitations would be required for a fully consistent quasiparticle description, and we have already noticed in section \ref{sec:o3} that infrared effects appear to be relevant at the next order in the expansion in $e$.

Finally, from the diagrammatic expansion of $\ln Z$ it is clear that the quasiparticle picture cannot describe all terms of higher orders in $e$. 
In FTQED the quasiparticle picture corresponds to using resummed thermal propagators (rather than free thermal propagators) in loop computations. This is, for instance,~commonly done in the so-called ``hard thermal loop'' resummation.
However, there are many higher-order contributions to $\ln Z$ that cannot be obtained from lower-order diagrams by simply replacing the bare propagators with resummed ones, e.g., all diagrams that involve vertex corrections.
This shortcoming clearly indicates that the 
quasiparticle picture cannot describe the
{\it bulk thermodynamic properties} of a plasma beyond the leading-order terms, even though the picture is extremely useful in the computation of {\it interaction rates} in cosmological settings, see,~e.g.,~\cite{Drewes:2010pf,Drewes:2013iaa}.
Indeed, it has been shown for both scalar field theories~\cite{Anisimov:2008dz} and QED~\cite{Leermakers:1985ii} at high temperatures
as well as in condensed matter systems~\cite{Suskov,Sandvik:2010hs}
that the stress-energy-momentum tensor of a thermal plasma can be split up into (i) a quasiparticle ideal gas component with temperature-dependent dispersion relations,  plus (ii)  an additional contribution that acts as a kind of ``interaction energy''.


\subsubsection{Is $N_{\rm eff}=3.052$?}

Central to the result of~\cite{Grohs2016} appears to be the claim that the FTQED corrections have been performed ``non-perturbatively''.  The term ``non-perturbatively'' together with the equations given in~\cite{Grohs2016}  suggests to us that they had adopted the same quasiparticle ideal gas description of~\cite{Heckler:1994tv}, 
but without Taylor-expanding  $P^{\rm qp}$ in powers of $e$ (``non-perturbative'').  Given the smallness of the fine structure constant~$\alpha$, this so-called ``non-perturbative'' procedure is tantamount to omitting the crucial $1/2$ factor that must be inserted before the ${\cal O}(e^2)$  pressure term~$P^{\rm qp}_{e^{2}}$.
This also means that following this (incorrect) recipe one must arrive at a finite-temperature correction to $N_{\rm eff}$ that is a factor of two larger than established results.%
\footnote{Reference~\cite{Pitrou:2018cgg} also pointed to the same missing factor of $1/2$  as the source of error in~\cite{Grohs2016}.}

We therefore conclude that the claimed FTQED contribution of $\delta N_{\rm eff} \simeq 0.02$ (and hence $N_{\rm eff}=3.052$)~\cite{Grohs2016} is incorrect, and most probably stemmed from a misinterpretation of the thermal mass/quasiparticle ideal gas narrative as laid down in~\cite{Heckler:1994tv}.%
\footnote{It is important to emphasise that the ${\cal O}(e^2)$ FTQED correction to the QED equation of state presented in~\cite{Heckler:1994tv} is itself correct, and in fact  follows from the same $\mathcal{O}(e^{2})$ pressure correction~(\ref{e^2_pressure}) used in this work.   Our criticism pertains only to the subsequent use by others of the thermal mass/quasiparticle ideal gas description~\cite{Heckler:1994tv}  as a defining property of finite-temperature effects.}


\section{Estimates of the neutrino decoupling temperature $T_d$  from real-time finite-temperature field theory}
\label{real_time_FTQFT}

What remains to be determined is the neutrino decoupling temperature $T_d$.  In the instantaneous decoupling limit, the decoupling temperature is defined via $\Gamma_\nu (T_d)=H(T_d)$, where $\Gamma_\nu (T)$ is the interaction rate per neutrino with the QED thermal bath from which it decouples, and $H(T)$ is the Hubble expansion rate.  In the timeframe of interest, the latter is given by
\begin{equation}
H^{2}(T) = \frac{8 \pi G}{3} \left[\rho (T)+ \rho_{\nu} (T)\right]
\end{equation}
where $G$ is the gravitational constant, $\rho$ is the energy density of the QED plasma discussed in detail in section~\ref{FTQED_intro}, and 
\begin{equation}
\rho_{\nu} = 3 \times \frac{7}{8} \frac{\pi^{2}}{30} g_{\nu} T_{\nu}^{4}
\end{equation}
is the energy density of the neutrino sector.  For the purpose of establishing the neutrino decoupling temperature, $T_{\nu}$ should be set to the plasma temperature~$T$.

We use the Schwinger--Keldysh formalism~\cite{Schwinger:1960qe,Bakshi:1962dv,Bakshi:1963bn,Keldysh:1964ud} of nonequilibrium quantum field theory to establish the neutrino interaction rate~$\Gamma_{\nu}$.
The Schwinger--Keldysh formalism (see, e.g.,~\cite{Chou:1984es,Berges:2004yj} for introductory articles) provides a convenient tool to obtain quantum kinetic equations at any desired order in $e$ and $G_F$. It has previously been applied to neutrino kinematics in dense and high-temperature environments~\cite{Vlasenko:2013fja,Blaschke:2016xxt}. In near-equilibrium situations the formalism practically reproduces the results of the real-time formalism of thermal field theory (see, e.g.~\cite{lebellac1996,Tututi:2002gz}).
Here, we work only at leading order in $e$, in which case the formalism  reproduces exactly the standard computation of the semi-classical Boltzmann collision integral, as  we shall illustrate in appendix~\ref{BoltzmannEquivalence}.  
While this clearly means that we are cracking nuts with a sledgehammer in the present computation, the approach we take here has the advantage that higher-order corrections can be systematically included 
 in follow-up works.

Using the Schwinger--Keldysh formalism it can be shown that the neutrino occupation numbers follow a generalised Boltzmann equation~\cite{Cirigliano:2009yt,Fidler:2011yq}, which in the absence of oscillations takes the form~\cite{Drewes:2012qw}
 \begin{equation}
\label{eq:gainlossboltzmann}
    \frac{\dd{ f_{\nu ,\mathbf{p}}}}{\dd{t}}= (1-f_{\nu,\mathbf{p}})\Gamma^<_\mathbf{p}-f_{\nu,\mathbf{p}}\Gamma^>_\mathbf{p},
\end{equation}
where
\begin{align}
\label{Gamma_definitions}
    \Gamma^{\gtrless}_\mathbf{p} &= \left.\frac{\mp 1}{ 2  p^0}{\rm Tr}[\slashed{p}\Pi^\gtrless]\right|_{p^0=\Omega_\mathbf{p}} 
    \end{align}
are the production ($<$; ``gain'') and destruction ($>$; ``loss'') terms, respectively, 
which are determined from the Wightman self-energies of opposite Schwinger--Keldysh polarities $\Pi^\lessgtr$.
Note that $ \Gamma^>_\mathbf{p}$ and $\Gamma^< _\mathbf{p}$ are functionals of all distribution functions in the plasma---including $f_{\nu ,\mathbf{p}}$,
and the master equation \eqref{eq:gainlossboltzmann} formally incorporates all contributions to the collision term at all orders in the couplings, provided that perturbation theory can be applied and the bulk properties of the plasma change only adiabatically~\cite{Drewes:2012qw}.

Since all species are in thermal equilibrium at $T>T_d$, we can use thermal propagators to compute $\Pi^\lessgtr$ so that the self-energies  satisfy the 
fermionic Kubo--Martin--Schwinger (KMS) relation
\begin{align}
\Pi^> = - e^{\Omega_\mathbf{p}/T}\Pi^<, \label{KMS}
\end{align}
in which $\Omega_\mathbf{p}$ is the full effective frequency dictating the dispersion relation of the neutrinos.
It then follows that the gain and loss terms~(\ref{Gamma_definitions}) obey the detailed balance condition $\Gamma^>_\mathbf{p}/\Gamma^< _\mathbf{p}= e^{\beta \Omega_\mathbf{p}}$, and the generalised Boltzmann equation~\eqref{eq:gainlossboltzmann} now simplifies to
\begin{equation}\label{Boltzmann-eq}
   \frac{\dd{f_{\nu,\mathbf{p}}}}{\dd{t}} = -\Gamma_\mathbf{p}\left[f_{\nu,\mathbf{p}} - f_D(\Omega_\mathbf{p}) \right], 
\end{equation}
where we can define a mode-dependent interaction rate, 
\begin{equation}
\begin{aligned}
\label{eq:master}
  \Gamma_\mathbf{p} & \equiv  \Gamma^>_\mathbf{p}+\Gamma^< _\mathbf{p} \\
  &=   \left.\frac{1}{2   p^0f_D(p^0)} {\rm Tr}\left[\slashed{p} \Pi^<\right]\right|_{p^0=\Omega_\mathbf{p}}.
  \end{aligned}
\end{equation}
Note that $\Gamma_\mathbf{p}$ can be related to the imaginary part of the retarded self-energy $\Pi^R$  of the neutrino at the (quasi)particle pole,
\begin{equation}
\label{eq:interactionrate}
    \Gamma_\mathbf{p}= - \left.\frac{1}{ p^0} {\rm Im Tr}\left[\slashed{p} \Pi^R\right]\right|_{p^0=\Omega_\mathbf{p}},
\end{equation}
following from the relation $2  {\rm Im}\Pi^R=  \Pi^> - \Pi^<$.  Indeed, the fact that a collision rate can be related to the imaginary part of a self-energy is simply a manifestation of the optical theorem and cutting rules at finite temperature~\cite{Weldon:1983jn}. See also \cite{Landshoff:1996ta,Gelis:1997zv,Bedaque:1996af}.

The form of equation~(\ref{Boltzmann-eq}) immediately suggests that it is the mode-dependent interaction rate  $\Gamma_\mathbf{p}$ that is responsible for driving the neutrino occupation number at mode~$\mathbf{p}$ back to its equilibrium value.   We therefore connect $\Gamma_\mathbf{p}$ to the interaction rate per neutrino $\Gamma_{\nu}$ required to compute the neutrino decoupling temperature (details to follow in section~\ref{sec:td}), and equation~(\ref{eq:master}) is our master equation for the rest of the analysis.


\subsection{The neutrino--QED plasma interaction rate from Fermi theory}

For the problem at hand, the particle interaction rate per particle, $\Gamma_{s}$, by definition counts only those weak-interaction processes that directly link the neutrino sector to the QED plasma.    Counting only $2 \to 2$ scattering processes, these are
\begin{equation}
\begin{aligned}
\label{eq:weakprocesses}
\nu_{e} \bar{\nu}_{e} &\to e^{+} e^{-}, \\
\nu_{e} e^{\pm} &\to \nu_{e}  e^{\pm},\\
\bar{\nu}_{e} e^{\pm} &\to \bar{\nu}_{e}  e^{\pm}.
\end{aligned}
\end{equation}
Note that we have neglected those processes that either (i)~do not involve an electron/positron, or (ii)~pertain to a muon or tau neutrino.  The former do not play a direct role in connecting the neutrino and the QED sector.  In the latter case, we assume the muon and tau neutrino populations to be kept in thermal equilibrium with the electron neutrino population through a combination of large-angle flavour oscillations and neutrino-neutrino scattering, which typically remain efficient beyond electron neutrino decoupling from the QED plasma~\cite{Dolgov:2002wy}.

\begin{figure}[t]
\centering
\includegraphics[width=15cm]{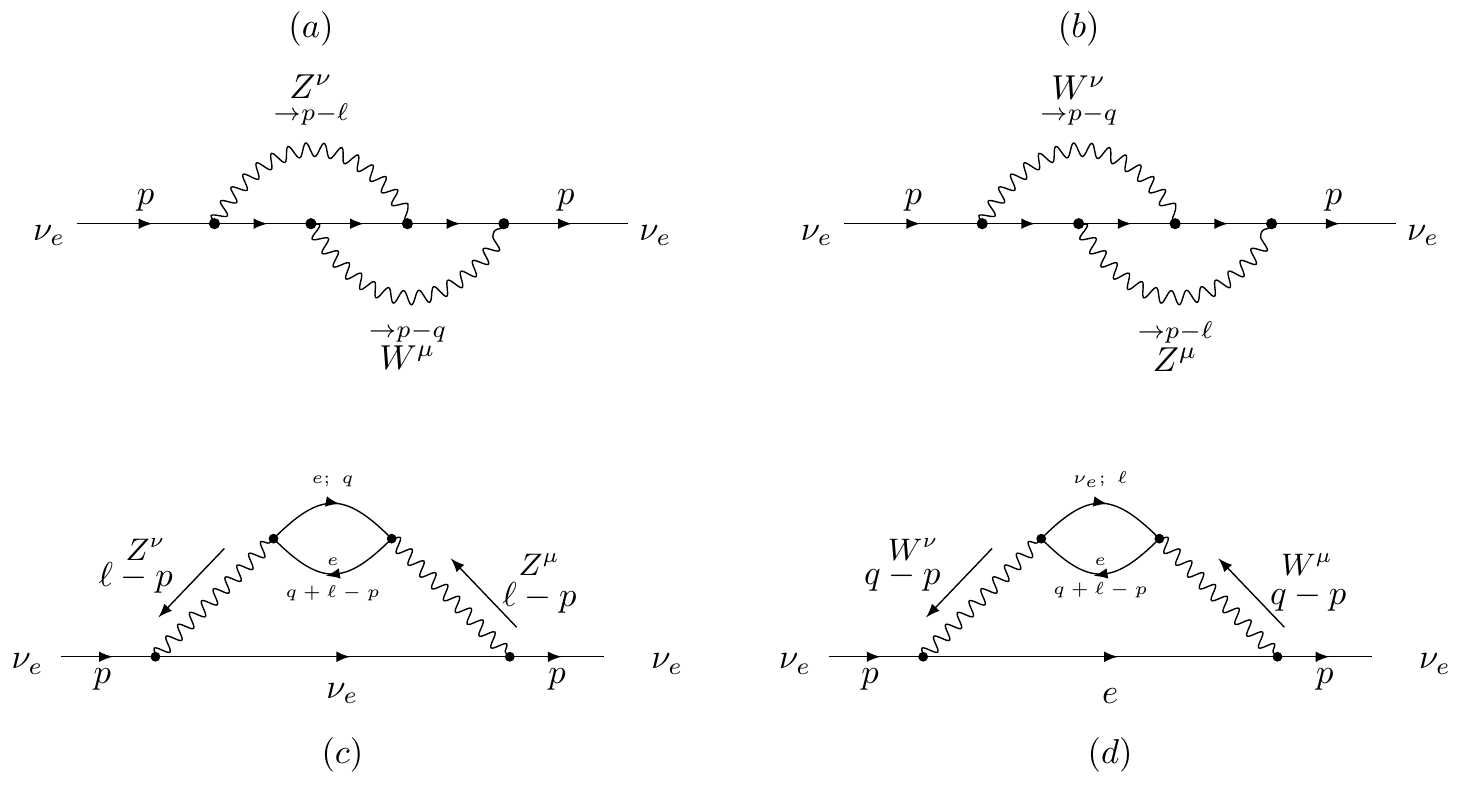}
 \caption{Leading-order weak self-energy diagrams that contain an imaginary component.  For notational simplicity  we have not labelled every internal fermion line and momentum.  The former is  however easy to deduce from the bosons connected to the vertices, and momentum is taken to flow in the same direction as the fermion number flow.}
 \label{selfenergy_for_weakrates}
\end{figure}

The self-energy diagrams encapsulating the scattering processes~(\ref{eq:weakprocesses}) are shown in figure~\ref{selfenergy_for_weakrates}.  Following 
SM Feynman rules~\cite{Romao:2012pq} and integrating out the massive gauge bosons in the Fermi limit, we find
\begin{equation}
\begin{aligned}
 {\rm Tr} \left[ \slashed{p} \Pi^{ba}_{(a)}(p)\right] = &-2  \left(\frac{G_F}{\sqrt{2}}\right)^2 \int \frac{\dd ^4 \ell \dd ^4 q}{(2\pi)^8}{\rm Tr}\Bigg[\slashed{p}\gamma^\mu (1-\gamma_5) i S^{ba}_e (q) \gamma^\nu (g_{V,e}-g_{A,e}\gamma_5) \\
    & \qquad \qquad \times iS^{ab}_e(\ell+ q -p)  \gamma_\mu (1 -  \gamma_5) i S^{ba}_\nu(\ell) \gamma_\nu (1-  \gamma_5)  \Bigg], \label{Kite1} 
    \end{aligned}
    \end{equation}
  \begin{equation}     
    \begin{aligned}
     \label{Kite2} 
   {\rm Tr} \left[ \slashed{p} \Pi^{ba}_{(b)}(p)\right] = & - 2  \left(\frac{G_F}{\sqrt{2}}\right)^2 \int \frac{\dd ^4 \ell \dd ^4 q}{(2\pi)^8}{\rm Tr}\Bigg[\slashed{p}\gamma^\mu (1-\gamma_5) i S^{ba}_\nu (\ell) \gamma^\nu (1-\gamma_5) \\
    & \qquad \qquad \times iS^{ab}_{e}(\ell+q-p)  \gamma_\mu (g_{V,e} - g_{A,e} \gamma_5) i S^{ba}_{e}(q) \gamma_\nu (1 - \gamma_5)  \Bigg], 
       \end{aligned}
      \end{equation}
   \begin{equation}     
    \begin{aligned}
     \label{Bubble1} 
   {\rm Tr} \left[ \slashed{p} \Pi^{ba}_{(c)}(p)\right] = & 4 \left(\frac{G_F}{\sqrt{2}}\right)^2 \int \frac{\dd ^4 \ell \dd ^4  q}{(2\pi)^8}{\rm Tr}\left[\slashed{p}\gamma^\mu (1-\gamma_5) i S^{ba}_\nu (\ell) \gamma^\nu (1-\gamma_5)\right] \\
    & \quad  \times  {\rm Tr}\left[\gamma_\mu (g_{V,e} - g_{A,e} \gamma_5) i S^{ba}_e(q) \gamma_\nu (g_{V,e} - g_{A,e} \gamma_5) i S^{ab}_e (\ell+ q-p) \right],\\
 \end{aligned}
 \end{equation}   
\begin{equation} 
   \begin{aligned}     
        \label{Bubble2} 
 {\rm Tr} \left[ \slashed{p} \Pi^{ba}_{(d)}(p)\right] = & \left(\frac{G_F}{\sqrt{2}}\right)^2 \int \frac{\dd ^4 \ell \dd ^4  q}{(2\pi)^8}{\rm Tr}\left[\slashed{p}\gamma^\mu (1-\gamma_5) i S^{ba}_e (q) \gamma^\nu (1-\gamma_5)\right]  \\
    & \qquad \qquad \times  {\rm Tr}\left[\gamma_\mu (1 -  \gamma_5) i S^{ba}_\nu(\ell) \gamma_\nu (1 -  \gamma_5) i S^{ab}_e(\ell+q-p) \right],
\end{aligned}
\end{equation}
where the superscripts $a$ and $b$ are real-time contour labels, $g_{V,e}= -\frac{1}{4} + \sin^2 \theta_W$, $g_{A,e}= -\frac{1}{4}$, and $g_{V,e}^\nu = g_{A,e}^\nu = \frac{1}{4}$. Note that we have chosen the loop momenta such that the internal neutrino always carries the momentum $\ell$ and the electrons  $q$ and $\ell+q-p$; this will ease the calculation later on. 
The total self-energy is simply a sum of these four contributions.

We are interested to compute $\Pi^<(p)$.  This corresponds to setting the contour indices to $a=-$ and $b=+$,  so that  $\Pi^{+-}(p)=\Pi^<(p)$, and only Wightman propagators $S^{\overset{+-}{-+}}_{\psi}=S^{\lessgtr}_{\psi}$, where $\psi= \nu_{e}, e$, appear in the expressions~(\ref{Kite1}) to (\ref{Bubble2}).  In momentum space, these read
\begin{equation}
\begin{aligned}
\label{eq:wightman}
   i S^>_{\psi}(p) & =(1-f_D(p^0))\rho_{ \psi}(p), \\
    i S^<_{\psi}(p) & =- f_D(p^0)\rho_{\psi} (p),
\end{aligned}
\end{equation}
where $\rho_{\psi} (p)$ is the spectral density.  For the present calculation we use the leading-order spectral density, i.e., the spectral density of a free fermion of mass $m_{\psi}$,
\begin{equation}
\label{eq:freefermionspectraldensity}
    \rho_{\psi, f}(p)=(2\pi) \textrm{sgn} (p^0) (\slashed{p}+m_{\psi}) \delta(p^2-m_{\psi}^2),
\end{equation}
which has only one sharp peak at $p^{2}=m_{\psi}^{2}$, i.e., the vacuum particle mass shell.  
Note however that the full resummed spectral density can have additional poles that correspond to collective plasma excitations in the quasiparticle picture, as well as continuous parts that encapsulate multiparticle states and higher order scattering processes (see~\cite{Drewes:2010pf,Drewes:2013iaa} for a discussion).
Using the approximation~(\ref{eq:freefermionspectraldensity}) also implies that we neglect the effect of thermal masses in the collision integral.

\begin{figure}[t]
    \centering
\includegraphics[width=12cm]{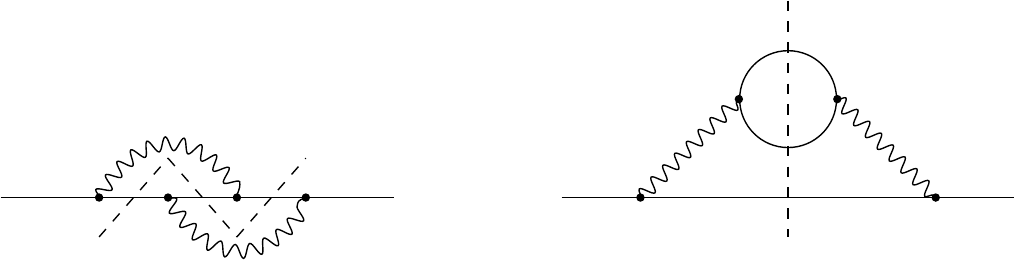}
    \caption{Finite temperature Cutkosky cuts for the Wightman self-energies depicted in figure~\ref{selfenergy_for_weakrates}.}
    \label{Cutkosky_Cuts}
\end{figure}

From here on, the connection of our approach to the usual kinetic theory method of calculating $\Gamma_{\nu}$ is clear: the free-fermion Wightman propagators can now be expressed as~\cite{Das:1997gg}
\begin{equation}
\label{eq:freefermionwightman}
   S_{\psi,f}^{\lessgtr}(p) = (f_D(|p^0|)-\theta (\mp p^0)  )(2\pi i) (\slashed{p}+m_\psi)\delta(p^2-m_\psi^2),
\end{equation}
where $\theta(x)$ denotes a Heaviside step function; these put the internal fermion lines of the self-energy diagrams on shell.  This amounts diagrammatically to cutting through all internal fermion lines, as shown in figure~\ref{Cutkosky_Cuts}, so that the resulting separate pieces are nothing but tree-level diagrams corresponding to the $2 \to 2$ scattering processes~(\ref{eq:weakprocesses}) and their complex conjugates.  We demonstrate in detail the correspondence between the two approaches in appendix~\ref{appendix_weak_rates}. Suffice it to say here that, while at leading order the two approaches are equivalent, nonequilibrium field theory methods clearly offer a more self-consistent way to incorporate higher-order effects.

\subsection{The neutrino decoupling temperature}
\label{sec:td}

Evaluating the self-energy contributions~(\ref{Kite1}) to (\ref{Bubble2}) assuming a free fermion-spectral density~(\ref{eq:freefermionspectraldensity}) for both $\psi=\nu_{e},e$, we find a mode-dependent interaction rate
\begin{equation}
\begin{aligned}
\label{weak_rate}
    \Gamma_{\mathbf{p}}(T) = &  \frac{(e^{|\mathbf{p}|/T}+1)}{2 |\mathbf{p|}} \sum_{i=a,b,c,d} \frac{\mathcal{C}_i \; G_F^2}{4(2\pi)^4} \iint_0^\infty \dd |\mathbf{q}| \dd |\boldsymbol{\ell}| \frac{|\mathbf{q}|^2 |\boldsymbol{\ell}|}{  E_q} \int_{-1}^{+1}\dd\cos\alpha \sum_{\epsilon,\tau = \pm 1}\left\{\frac{ \pi \theta(\tilde{b}^2-4\tilde{a}\tilde{c})}{\sqrt{|\tilde{a}|}} \right. \\
  & \times  \Big[f_D(|\ell^0 + q^0 -p^0|)-\theta (\ell^0 + q^0 -p^0)  \Big]\Big[f_D(|\ell^0|)-\theta (- \ell^0)  \Big]\Big[f_D(|q^0|)-\theta (-q^0)  \Big] \\
    &\left. \qquad \qquad \times\left. \sum_{(mn)}  A^i_{(mn)}\left[G^0_{(mn)} + \frac{\tilde{b}}{2|\tilde{a}|} G^1_{(mn)} + \left(\frac{3 \tilde{b}^2 + 4 \tilde{c} | \tilde{a} |}{8 \tilde{a}^2}\right) G^2_{(mn)}\right]\right\}
\right|_{\overset{ q^0=\epsilon E_q}{\ell^0=\tau |\boldsymbol{\ell}|}}.
\end{aligned}
\end{equation}
The definitions of the coefficients $\mathcal{C}_i$, $A^i_{(mn)}$, and $G^i_{(mn)}$ can be found in tables~\ref{tab:coeff1} and \ref{tab:coeff2}, while $\tilde{a}$, $\tilde{b}$, and $\tilde{c}$ are $\epsilon$- and $\tau$- dependent functions of  $|\mathbf{p}|$, $|\mathbf{q}|$,  $|\boldsymbol{\ell}|$, and $\cos \alpha$ given in equations~(\ref{eq:a}) to (\ref{eq:c}).
We detail the  full calculation of $\Gamma_{\mathbf{p}}(T)$ in appendix~\ref{weak_rate_calculation}.

\begin{figure}[t]
    \centering
\includegraphics[height=13cm,angle=270]{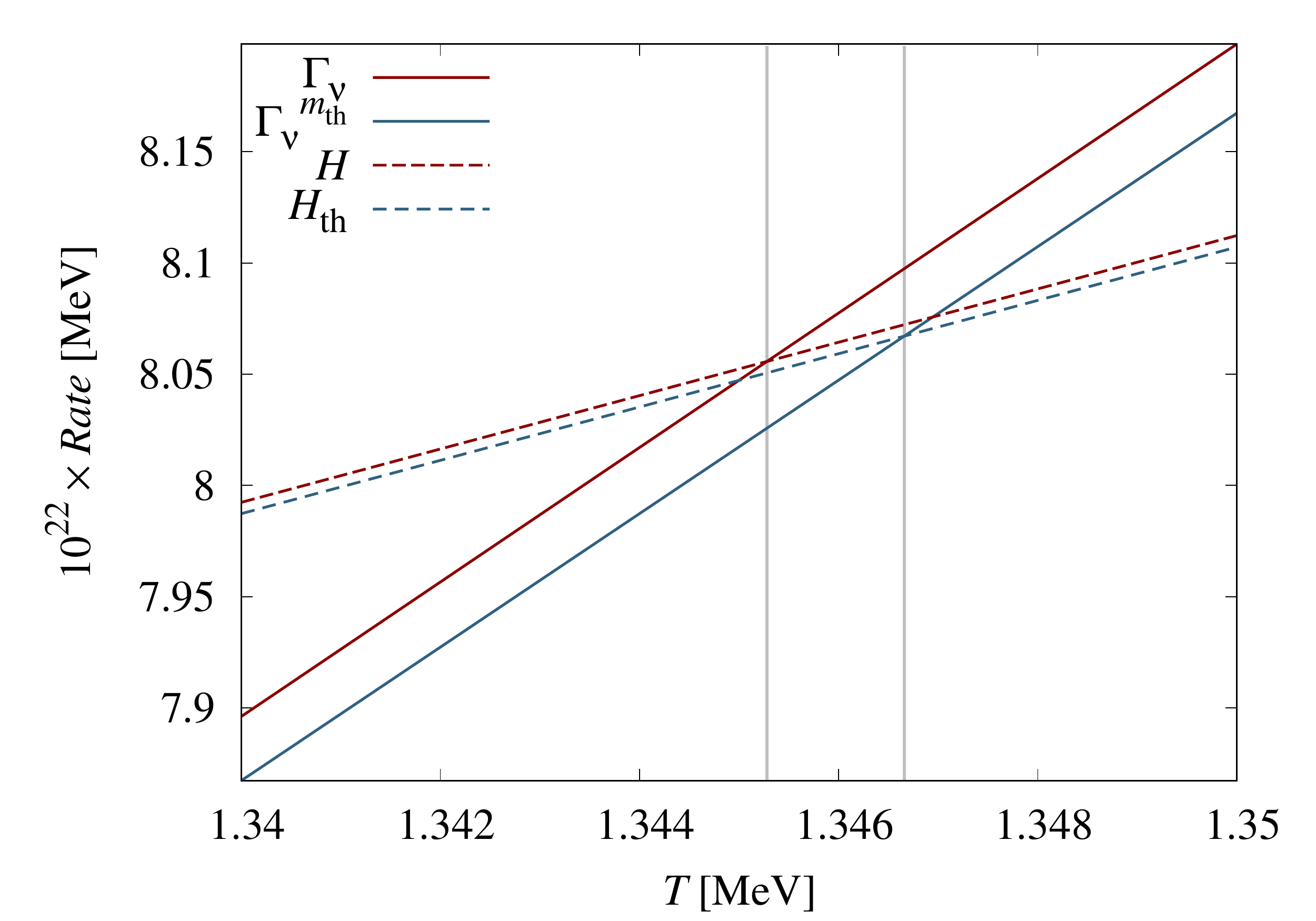}
    \caption{Interaction rate per neutrino with the QED plasma $\Gamma_\nu$ (solid lines) versus the Hubble expansion rate $H$ (dashed lines) as functions of the temperature of the QED plasma~$T$.
Rates without FTQED corrections are indicated in red, while those with FTQED corrections are in blue.}
    \label{GammaH}
\end{figure}

To connect the mode-dependent $\Gamma_{\mathbf{p}}$ to the interaction rate per neutrino $\Gamma_{\nu}$, we identify $\Gamma_{\nu}$ with the 
 $\Gamma_{\mathbf{p}}$ evaluated at some representative neutrino momentum, 
 \begin{equation}
 \Gamma_{\nu} \equiv\left.  \Gamma_{\mathbf{p}} \right|_{|\mathbf{p}|=\langle |\mathbf{p}| \rangle},
 \end{equation}
 which we choose here to be the mean momentum  $\langle |\mathbf{p}| \rangle\simeq 3.15 T$.  (Alternatively, we can identify $\Gamma_{\nu}(T)$ with the momentum-averaged $\langle \Gamma_{\mathbf{p}}\rangle(T)$, weighted by the Fermi--Dirac distribution function; this procedure is however computationally more expensive, and therefore not used here.)     Figure~\ref{GammaH} shows the interaction rate $\Gamma_\nu(T)$ thus defined as a function of the QED plasma temperature~$T$, alongside the Hubble expansion rate $H(T)$  assuming the QED plasma to be an ideal gas.

Solving $ \Gamma_\nu(T_{d}) - H(T_{d})=0$ for $T_{d}$, we find
\begin{equation}
    T_d= 1.3453~\rm{MeV}.
\end{equation}
This is the neutrino decoupling temperature used throughout this work, and is 
consistent with the findings of~\cite{Fornengo:1997wa} to about $5\%$.  We attribute the difference primarily to the different definitions of $\Gamma_\nu$: while we have defined in this work $\Gamma_\nu$ as the mode-dependent rate $\Gamma_\mathbf{p}$ evaluated at $|\mathbf{p}|=\langle |\mathbf{p}| \rangle\simeq 3.15 T$, the definition used in~\cite{Fornengo:1997wa} is essentially equivalent to identifying $\Gamma_\nu$ with the momentum-averaged destruction rate $\langle \Gamma_\mathbf{p}^> \rangle$.

\subsubsection{FTQED corrections to the decoupling temperature}

An immediate and straightforward way to incorporate some form of FTQED corrections in the estimate of~$T_d$ is to replace in the interaction rate~(\ref{weak_rate}) all occurrences of the vacuum electron mass $m_{e}^{2}$ by its thermally-corrected counterpart $m_{e}^{2}+ \delta m_{e}^{2}(p,T)$ given in equation~(\ref{eq:meth}).   This is essentially equivalent to using $\mathcal{O}(e^2)$ resummed electron propagators in the calculation of the neutrino self-energy, and is especially simple to compute if we neglect the momentum-dependent part of $\delta m_{e}^{2}(p,T)$, i.e., the third, logarithmic term of equation~(\ref{eq:meth}), which is in any case known to be subdominant~\cite{Mangano:2001iu}. Care must also be taken to incorporate the higher-order energy density corrections $\rho^{(n)}$ previously calculated in section~\ref{FTQED_intro} into the Hubble expansion rate, and not simply alter the ideal gas phase space distribution by replacing the vacuum electron mass  with its thermally-corrected counterpart.\footnote{
At first glance it might seem inconsistent that we freely replace vacuum masses with thermal masses in order to 
 estimate FTQED corrections to the weak interaction rate~(\ref{weak_rate}), while we argue in section~\ref{thermal_mass_interpretation}
that this procedure is not consistent when it comes to the equation of state (and thus the Hubble rate).
The key point is that the quasiparticle picture offers a suitable scheme to compute interaction rates in resummed perturbation theory (see, e.g.,~\cite{Drewes:2010pf}), but fails to accurately describe the bulk properties of the plasma.
.
}

Figure~\ref{GammaH} shows the thermal-mass-corrected interaction rate $\Gamma_\nu^{m_{\rm th}}(T)$ as a function of the QED plasma temperature, together with the ${\cal O}(e^2)$ thermally-corrected Hubble expansion rate $H_{\rm th}(T)$.
Solving $\Gamma_\nu^{m_{\rm th}}(T_{d}^{m_{\rm th}})-H_{\rm th}(T_{d}^{m_{\rm th}})=0$ for the thermal-mass-corrected neutrino decoupling temperature $T_{d}^{m_{\rm th}}$, we find
\begin{equation}
    T_d^{m_{\rm th}} = 1.3467~{\rm MeV},
\end{equation}
a deviation of $0.1\%$ from the uncorrected $T_{d}$, consistent with the findings of~\cite{Fornengo:1997wa}.%
\footnote{Note that~\cite{Fornengo:1997wa} always assumed ideal gases  when evaluating $H(T)$. Their thermally-corrected neutrino decoupling temperature therefore corresponds to the intersection of the red dashed and blue solid lines in figure~\ref{GammaH}, which generally returns a higher value of $T_d$ than the intersection of the two blue lines.}
Reading off figures~\ref{fig:Allresults} and~\ref{fig:Allresultsthermal}, this deviation essentially shifts the final $N_{\rm eff}$ by an amount
\begin{equation}
\delta N_{\rm eff}^{m_{\rm th}} \simeq -0.000080,
\end{equation}
 larger than the ${\cal O}(e^2)$ logarithmic correction to the QED equation of state discussed in section~\ref{sec:o2log}, but still an order of magnitude smaller than the ${\cal O}(e^3)$ correction of section~\ref{sec:o3}.  Interestingly, the most recent precision  calculations of $N_{\rm eff}$ do incorporate thermal-mass-corrected weak rates in their modelling of 
 neutrino energy transport around decoupling~\cite{deSalas,Gariazzo:2019gyi}.  What we have demonstrated here is that, on its own, this correction has no significant impact on $N_{\rm eff}$.

Of course, a fully self-consistent treatment of ${\cal O}(e^{2})$ FTQED corrections to the weak rates goes beyond using resummed electron propagators: corrections to the weak vertices in the self-energy diagrams of figure~\ref{selfenergy_for_weakrates}, for example, also enter at ${\cal O}(e^{2})$, which would in principle shift $N_{\rm eff}$ again.  A complete treatment of  
${\cal O}(e^{2})$ FTQED corrections to the weak rates will be presented in a future publication.


\section{Conclusions}
\label{sec:conclusions}


\begin{table}[t]
\centering
\begin{tabular}{|c|cc|c|}
\hline
$x$ & $\delta N_{\rm eff}^{x}$ ($T_{d}=1.3453$~MeV) &   $\delta N_{\rm eff}^{x}$ ($T^{m_{\rm th}}_{d}=1.46$~MeV) & Included in~\cite{deSalas,Gariazzo:2019gyi} \\
\hline
\hline
& \multicolumn{2}{c|}{QED equation of state corrections}& \\
\hline
$\slashed{\rm Rel}$ & 0.039895 & 0.033903  & Yes\\
$(2) \slashed{\ln}$ &  0.010121 & 0.010173 & Yes \\
$(2)\ln$ & $-0.000050$ & $-0.000043$ & No\\
$(3)$ & $- 0.000952$ & $-0.000951$ & No\\
$(4)$ & $\simeq 3.5 \times 10^{-6}$ & $\simeq 3.5 \times 10^{-6}$  & No\\
\hline
\hline
& \multicolumn{2}{c|}{Weak rate corrections}& \\
\hline
$m_{\rm th}$ &$-0.000080$ & $-0.000067$ & Yes  \\
\hline
\hline
{\bf Total }& $\mathbf{0.048937}$ & $\mathbf{0.043019}$ &\\
\hline
\end{tabular}
\caption{Summary of the various SM contributions to $N_{\rm eff}$ considered in this work.  The (instantaneous) neutrino decoupling temperature $T_{d}=  1.3453$~MeV is that established in section~\ref{real_time_FTQFT}, while $T^{m_{\rm th}}_{d}=1.46$~MeV has been obtained from matching  $\delta N_{\rm eff}^{\slashed{\rm Rel}+(2)\slashed{\ln}}+\delta N_{\rm eff}^{m_{\rm th}}$ to $0.044$~\cite{Gariazzo:2019gyi}.   The last column indicates whether or not the correction has been implemented in the most recent full neutrino energy transport calculations of~\cite{deSalas,Gariazzo:2019gyi}.\label{tab:summary}}

\end{table}


We have examined and quantified precisely in this work several aspects of Standard Model physics that drive the theoretical value of the effective number of neutrinos $N_{\rm eff}$ away from~3.  The expected deviations are summarised in table~\ref{tab:summary}.

The dominant departure from $N_{\rm eff}=3$ comes from dropping the assumption that the QED plasma is ultra-relativistic at the time of neutrino decoupling, i.e., what we have dubbed the ``ultra-relativistic'' (Rel) approximation, as it is equivalent to sending $T_{d}/m_{e} \to \infty$.  A finite $T_{d}/m_{e}$ enables some energy from $e^\pm$-annihilation 
to be transferred to the neutrinos when the neutrino and QED sectors are in thermal equilibrium, and, depending on the precise~$T_{d}$, can incur a change in $N_{\rm eff}$,  $\delta N_{\rm eff}^{\slashed{\rm Rel}}$, as large as a few percent.  This effect is
 often conflated with non-instantaneous decoupling in the literature, the latter of which can induce out-of-equilibrium energy transfer and distort the equilibrium distributions.   We emphasise here that the two phenomena are distinct, and genuine  out-of-equilibrium effects on $N_{\rm eff}$ are in fact subdominant.

We then examined finite-temperature corrections to the QED equation of state and their effects on the $N_{\rm eff}$ parameter.  Specifically, beginning with the FTQED partition function up to and including ${\cal O}(e^4)$, we have quantified the role of each contribution to driving $N_{\rm eff}$ from~3.
 At leading-order ${\cal O}(e^2)$, we have been able to recover the established result $\delta N_{\rm eff}^{(2)\slashed{\ln}} \simeq 0.01$, and, in doing so, identify the source of error in~\cite{Grohs2016} (which purportedly included the same FTQED correction but obtained a correction twice as large as the establishment).  We further assessed a hitherto-neglected ${\cal O}(e^2)$ logarithmic contribution, and found its effect on~$N_{\rm eff}$, $\delta N_{\rm eff}^{(2)\ln} \simeq - 5 \times 10^{{-5}}$, to be even smaller than previously envisaged.  Relative to our four-significant-digit accuracy goal, this ${\cal O}(e^2)$ logarithmic correction---along with the  ${\cal O}(e^4)$ correction,  which contribution $\delta N_{\rm eff}^{(4)} \simeq 3 \times 10^{{-6}}$---can be considered optional to unnecessary in future calculations.
 
Of particular note, however, is the ${\cal O}(e^3)$ correction to the QED equation of state.  Originating from Debye screening, this correction contributes $\delta N_{\rm eff}^{(3)} \simeq 0.001$
across a broad range of plausible neutrino decoupling temperatures.  This is a sizeable correction not only relative to our accuracy goal, but also relative to the typical changes incurred in $N_{\rm eff}$ between including and excluding neutrino oscillations in the modelling of out-of-equilibrium neutrino energy transport (up to $\delta N_{\rm eff}^{\rm osc} \simeq -0.0004$, depending on the transport code).  Thus, the 
${\cal O}(e^3)$ correction to the QED equation of state should be considered a necessary input in the precision computation of $N_{\rm eff}$.

Lastly, we re-estimated the neutrino decoupling temperature $T_{d}$ using nonequilibrium quantum field theory techniques, which have the advantage of providing a consistent and systematic framework for  the inclusion of FTQED corrections to the weak rates. The exercise returns at leading order $T_d=1.3453$~MeV, a result consistent to the estimate of~\cite{Fornengo:1997wa} to 5\% despite subtle differences in the definition of the decoupling temperature.   Correcting the weak rate calculations with ${\cal O}(e^{2})$ resummed electron propagators we found $T_{d}^{m_{\rm th}} = 1.3467$~MeV, corresponding to a  minute change in $N_{\rm eff}$ of $\delta N_{\rm eff}^{m_{\rm th}} \simeq  -8 \times 10^{-5}$.

A complete assessment of the various effects considered in this work on the final value of $N_{\rm eff}$ will necessitate an account of neutrino energy transport beyond the instantaneous decoupling approximation in the manner of~\cite{deSalas,Gariazzo:2019gyi}. 
  However, by  matching known corrections $\delta N_{\rm eff}^{\slashed{\rm Rel}+(2)\slashed{\ln}}+\delta N_{\rm eff}^{m_{\rm th}}$ to the result obtained in the most recent such calculation~\cite{Gariazzo:2019gyi} (see column 3 of table~\ref{tab:summary}), we deduce that, relative to $N_{\rm eff}  = 3.044$~\cite{Gariazzo:2019gyi},  the new effects found in this work should lower the number to $N_{\rm eff} = 3.043$.  The confirmation of this number via a full transport calculation will be presented in a future publication.

\acknowledgments 

JJB and Y$^3$W are supported in part by the Australian Government through the Australian Research Council's Discovery Project (project DP170102382) and Future Fellowship (project FT180100031) funding schemes. GB acknowledges the support of the National Fund for Scientific Research (F.R.S.- FNRS Beligum) through a FRIA grant.   MaD and
Y$^3$W  acknowledge support from the ASEM-DUO fellowship programme of the Belgian Acad\'emie de recherche et d'enseignement sup\'erieur (ARES).
Computational resources have been provided by the F.R.S.-FNRS Consortium des \'Equipements de Calcul Intensif (C\'ECI), funded by the Grant 2.5020.11 and by the Walloon Region.
We thank Dimi Culcer, Pablo de Salas, Stefano Gariazzo, Sergio Pastor, Michael Schmidt, and Oleg Sushkov for useful discussions.


\appendix

\section{Connection of the damping rate to the Boltzmann collision term}\label{appendix_weak_rates}\label{BoltzmannEquivalence}

We demonstrate in this section the correspondence, at leading order, between the neutrino damping rate calculated from the retarded self-energy represented by figure~\ref{selfenergy_for_weakrates}, and the usual Boltzmann collision term for the $2\to 2$ scattering processes~(\ref{eq:weakprocesses}).

For brevity we illustrate in detail only the self-energy diagram $(b)$ in figure~\ref{selfenergy_for_weakrates}, or, equivalently, equation~(\ref{Kite2}); the correspondence between the two approaches for the other three self-energy terms can be easily established in the same manner. For $a=-$ and $b=+$ (hence $b a \equiv <$ and $ a b \equiv >$) and plugging into equation~(\ref{Kite2}) the Wightman propagators~(\ref{eq:wightman}),
we have
\begin{equation}
\begin{aligned}
\label{eq:diagramb}
  {\rm Tr} \left[ \slashed{p} \Pi^<_{(b)}(p)\right]  =& - 2  \left(\frac{G_F}{\sqrt{2}}\right)^2 \int \frac{\dd ^4 \ell \dd ^4q}{(2\pi)^8}f_D(\ell^0)(1-f_D(\ell^0+q^0-p^0))f_D(q^0)  \\
     & \hspace{-20mm} \times {\rm Tr}\Bigg[\slashed{p}\gamma^\mu (1-\gamma_5) \rho_{\nu} (\ell) \gamma^\nu (1-\gamma_5)\rho_{e}(\ell+q-p)  \gamma_\mu (g_{V,e} - g_{A,e} \gamma_5) \rho_{e}(q) \gamma_\nu (1 - \gamma_5)  \Bigg],
\end{aligned}
\end{equation}
which is generally valid for any spectral density~$\rho_{\psi}(p)$.
At leading order we use the free spectral density~(\ref{eq:freefermionspectraldensity}), so that equation~(\ref{eq:diagramb}) reduces to
\begin{equation}
\begin{aligned}
    {\rm Tr} \left[ \slashed{p} \Pi^<_{(b)}(p)\right] =& -G_F^2 \int \frac{\dd ^4 \ell \dd ^4q}{(2\pi)^5}f_D(\ell^0)(1-f_D(\ell^0+q^0-p^0))f_D(q^0)  \\
    & \hspace{-10mm} \times {\rm sgn}(\ell^0){\rm sgn}(\ell^0+q^0-p^0){\rm sgn}(q^0) \delta(\ell^2) \delta((\ell+q-p)^2-m_{e}^2) \delta(q^2-m_{e}^2)   {\rm Tr}_b,
\end{aligned}
\end{equation}
where we have defined
\begin{equation}
{\rm Tr}_b\equiv{\rm Tr}\Big[\slashed{p}\gamma^\mu (1-\gamma_5) \slashed{\ell} \gamma^\nu (1-\gamma_5) (\slashed{\ell}+\slashed{q}-\slashed{p}+m_{e}) \gamma_\mu (g_{V,e} - g_{A,e} \gamma_5) (\slashed{q}+m_{e}) \gamma_\nu (1 - \gamma_5)\Big],
\end{equation}
which contains all of the Dirac algebra of the problem at hand.
Using resummed spectral densities in \eqref{eq:diagramb} would correspond to dressing all particles into quasiparticles with thermal masses and widths, which is a convenient way to introduce some higher order FTQED corrections, in particular (but not only) those assigned to thermal masses \cite{Drewes:2010pf}.

 In order to connect to kinetic theory and the Boltzmann collision term, we introduce  a $4$-dimensional Dirac delta distribution in $u$ and then integrate over $u$:
 \begin{equation}
\begin{aligned}
    {\rm Tr} \left[ \slashed{p} \Pi^<_{(b)}(p)\right] = & -G_F^2 \int \frac{\dd ^4 \ell \dd ^4q \dd^4 u}{(2\pi)^9}(2\pi)^4 \delta^{(4)}(u- q - \ell + p ) f_D(\ell^0)(1-f_D(u^0))f_D(q^0) \\
      & \qquad \times \;{\rm sgn}(\ell^0){\rm sgn}(u^0){\rm sgn}(q^0) \delta(\ell^2) \delta(u^2-m_{e}^2) \delta(q^2-m_{e}^2)   {\rm Tr}_b.
\end{aligned}
\end{equation}
Note that this  step does not change the physics of the expression; it merely recasts the expression in a more convenient form for our purposes.
Then, cycling through all possible sign combinations of $\ell^{0},u^{0}$, and~$q^{0}$ yields
\begin{equation}
\begin{aligned}
  {\rm Tr} \left[ \slashed{p} \Pi^<_{(b)}(p)\right] =&  -G_F^2 \int \frac{\dd ^3\ensuremath{\boldsymbol\ell} \dd ^3 \mathbf{q}  \dd ^3 \mathbf{u}}{(2\pi)^9 8 E_\ell E_q E_u}(2\pi)^4 \delta^{(3)}(\mathbf{u }-  \mathbf{q} - \boldsymbol\ell + \mathbf{p} )  \\
      & \times \Big[ f_\ell (1-f_u)f_q \delta(E_u-E_q-E_\ell + p^0){\rm Tr}_b^{+++}  \\
      & \qquad - f_\ell (1-f_u)(1-f_q) \delta(E_u+E_q-E_\ell + p^0){\rm Tr}_b^{++-}  \\
      &\qquad - f_\ell f_u f_q \delta(-E_u-E_q-E_\ell + p^0){\rm Tr}_b^{+-+} \\
      & \qquad - (1-f_\ell) (1-f_u)f_q \delta(E_u-E_q+E_\ell + p^0){\rm Tr}_b^{-++}  \\
      & \qquad + f_\ell f_u(1- f_q) \delta(-E_u+E_q-E_\ell + p^0){\rm Tr}_b^{+--}  \\
      & \qquad + (1-f_\ell) f_uf_q \delta(-E_u-E_q+E_\ell + p^0){\rm Tr}_b^{--+}  \\
      & \qquad+(1-f_\ell)(1- f_u)(1-f_q) \delta(E_u+E_q+E_\ell + p^0){\rm Tr}_b^{-+-}  \\
      & \qquad-(1-f_\ell) f_u(1-f_q) \delta(-E_u+E_q+E_\ell + p^0){\rm Tr}_b^{---}\Big],
       \label{b_collision_term}
\end{aligned}
\end{equation}
where we have defined
\begin{equation}
    {\rm Tr}_b^{ijk}\equiv \left.{\rm Tr}_b\right|_{\ell^0= i E_\ell, \; u^0=j E_u ,\; q^0=k E_q},
\end{equation}
and used the vacuum dispersion relations $E_\ell = |\boldsymbol{\ell}|$, $E_u^2=|\mathbf{u}|^2 + m_{e}^2$, and $E_q^2 \equiv |\mathbf{q}|^2 + m_{e}^2$.  The particle phase space distributions are labelled by their 4-momenta in the subscript, 
$f_{x} \equiv f_D(E_{x})$, and we have used the relation $f_D(-E)=1 - f_D(E)$.

\begin{figure}[t]
    \centering
 \includegraphics[width=15cm]{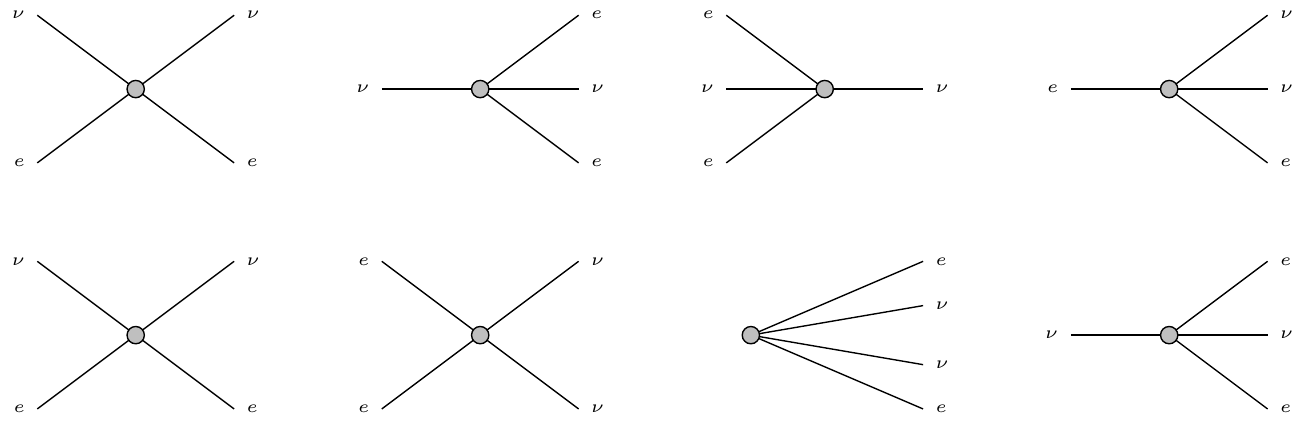}
     \caption{Scattering processes corresponding to the eight terms in equation~\eqref{b_collision_term}, read from left to right, top to bottom.  Only the $2 \to 2$ processes are kinematically allowed.}
    \label{Collision_Diagrams}
\end{figure}

In the form~(\ref{b_collision_term}), the physics of the self-energy is immediately discernible: the processes corresponding to each term can be identified by their initial and final state particles.   Specifically, initial state phase space distributions always appear simply as a factor~$f_{x}$, while final state phase space distributions come in the form of a Pauli-blocking factor~$(1-f_{x})$.  Furthermore, because we are computing ${\rm Tr} \left[ \slashed{p} \Pi^<_{(b)}(p)\right]$, which is associated with the production rate $\Gamma^{<}$ and hence accompanied by a Pauli-blocking factor $(1-f_{p})$ in the generalised Boltzmann equation~(\ref{eq:gainlossboltzmann}), the neutrino that carries the 4-momentum $p$ should also be interpreted as a final state particle.  Then, the eight processes corresponding to the eight terms of equation~(\ref{b_collision_term}) are simply those  shown in figure~\ref{Collision_Diagrams} (read from left to right, top to bottom).

Clearly, not all of the processes of  figure~\ref{Collision_Diagrams} are allowed.
Throwing away those processes that are kinematically forbidden, we are left with
\begin{equation}
\begin{aligned}
 {\rm Tr} \left[ \slashed{p} \Pi^<_{(b)}(p)\right] =&  -G_F^2 \int \frac{\dd ^3\ensuremath{\boldsymbol\ell} \dd ^3 \mathbf{q}  \dd ^3 \mathbf{u}}{(2\pi)^9 8 E_\ell E_q E_u}(2\pi)^4 \delta^{(3)}(\mathbf{u }-  \mathbf{q} - \boldsymbol\ell + \mathbf{p} ) \\
      & \times \Big[ f_\ell (1-f_u)f_q \delta(E_u-E_q-E_\ell + p^0){\rm Tr}_b^{+++}  \\
      & \qquad + f_\ell f_u(1- f_q) \delta(-E_u+E_q-E_\ell + p^0){\rm Tr}_b^{+--}  \\
      & \qquad + (1-f_\ell) f_uf_q \delta(-E_u-E_q+E_\ell + p^0){\rm Tr}_b^{--+}\Big].
       \label{b_collision_term_surviving}
\end{aligned}
\end{equation}
Repeating the calculation for the remaining three self-energy contributions~(\ref{Kite1}), (\ref{Bubble1}), and (\ref{Bubble2}), we always find the same phase space structure, the differences being entirely contained in the trace terms and the coupling constant.  Then, collecting all contributions, the total  mode-dependent production rate can now be written as
\begin{equation}
\begin{aligned}
    \Gamma^<_\mathbf{p}=&\; 
    \frac{ 1}{2 p^0}{\rm Tr} \left[\sum_{i'=a,b,c,d} \slashed{p} \Pi^<_{(i')}(p)\right] \\
    =&\;  -\frac{G_F^2}{2p^0} \int \frac{ \dd ^3\ensuremath{\boldsymbol\ell} \dd ^3 \mathbf{q} \dd ^3 \mathbf{u}}{(2\pi)^9 8 E_\ell E_q E_u}(2\pi)^4 \delta^{(3)}(\mathbf{u }-  \mathbf{q} - \boldsymbol\ell + \mathbf{p} )  \\
      & \qquad \times \Big[ f_\ell (1-f_u)f_q \delta(E_u-E_q-E_\ell + p^0){\rm Tr}^{+++}  \\
      & \qquad \qquad + f_\ell f_u(1- f_q) \delta(-E_u+E_q-E_\ell + p^0){\rm Tr}^{+--}  \\
      & \qquad \qquad + (1-f_\ell) f_uf_q \delta(-E_u-E_q+E_\ell + p^0){\rm Tr}^{--+}\Big],
      \label{Gamma_lessthan}
\end{aligned}
\end{equation}
where ${\rm Tr}^{ijk}\equiv \sum_{i'=a,b,c,d}\lambda_{i'}{\rm Tr}_{i'}^{ijk}$, with coefficients $\lambda_a=\lambda_b=1$, $\lambda_c=- 2$, and $\lambda_d=-1/2$.  The traces have been evaluated to be
\begin{equation}
\begin{aligned}
    {\rm Tr}_a &  = {\rm Tr}\Big[\slashed{p}\gamma^\mu (1-\gamma_5)( \slashed{q}+m_{e}) \gamma^\nu (g_{V,e}-g_{A,e}\gamma_5) (\slashed{u}+m_{e}) \gamma_\mu (1 -  \gamma_5) \slashed{\ell}\gamma_\nu (g_{V,e} - g_{A,e}\gamma_5)  \Big]  \\
     & = -128(g_{V,e}+g_{A,e})(p\cdot u)(\ell \cdot q) + 64 m_{e}^2 (g_{V,e}-g_{A,e})(p\cdot \ell), \label{eq:tra}
\end{aligned}
\end{equation}    
\begin{equation}
   \begin{aligned}  
          {\rm Tr}_b &  = {\rm Tr}\Big[\slashed{p}\gamma^\mu (1-\gamma_5) \slashed{\ell} \gamma^\nu (1-\gamma_5) (\slashed{u}+m_{e}) \gamma_\mu (g_{V,e} - g_{A,e} \gamma_5) (\slashed{q}+m_{e}) \gamma_\nu (1 - \gamma_5)  \Big]  \\
     & = -128(g_{V,e}+g_{A,e})(p\cdot u)(\ell \cdot q) + 64 m_{e}^2 (g_{V,e}-g_{A,e})(p\cdot \ell)={\rm Tr}_a,
\end{aligned}
\end{equation}
\begin{equation}
\begin{aligned}   
          {\rm Tr}_c &  = {\rm Tr}\Big[\slashed{p}\gamma^\mu (1-\gamma_5) \slashed{\ell} \gamma^\nu (1-\gamma_5)\Big]{\rm Tr}\Big[\gamma_\mu (g_{V,e}-g_{A,e}\gamma_5) (\slashed{q}+m_{e}) \gamma_\nu (g_{V,e} - g_{A,e} \gamma_5) (\slashed{u}+m_{e})  \Big]  \\
     & = 64(g_{V,e}^2+g_{A,e}^2)\Big((p\cdot q) (\ell \cdot u) + (p\cdot u)(\ell \cdot q)\Big) - 64 m_{e}^2 (g_{V,e}^2-g_{A,e}^2)(p\cdot \ell) \\
     & \qquad \qquad - 128 g_{V,e}g_{A,e}\Big((p\cdot q)(\ell \cdot u)-(p\cdot u)(\ell \cdot q)\Big),
\end{aligned}
\end{equation}     
\begin{equation}
\begin{aligned}     
     {\rm Tr}_d &  = {\rm Tr}\Big[\slashed{p}\gamma^\mu (1-\gamma_5) (\slashed{q}+m_{e}) \gamma^\nu (1-\gamma_5)\Big]{\rm Tr}\Big[\gamma_\mu (1-\gamma_5) \slashed{\ell} \gamma_\nu (1- \gamma_5) (\slashed{u}+m_{e}) \Big]  \\
     & = 256 (p\cdot u)(\ell \cdot q), \label{eq:trd}
\end{aligned}
\end{equation}
from which we conclude that the self-energy diagrams $(a)$ and $(b)$ contribute identically to the interaction rate. 

The corresponding mode-dependent destruction rate~$\Gamma_{\mathbf{p}}^{>}$ can be easily deduced from $\Gamma_{\mathbf{p}}^{<}$ by~(i) noting that, to turn  ${\rm Tr} \left[ \slashed{p} \Pi^<(p)\right]$ into  ${\rm Tr} \left[ \slashed{p} \Pi^>(p)\right]$, we need simply to replace the Wightman propagators $S_{\psi}^{\gtrless} \to S_{\psi}^{\lessgtr}$, which, through equation~(\ref{eq:wightman}), amounts to swapping $(1-f_x)\leftrightarrow -f_x$, and (ii) introducing an overall sign flip via the definition~(\ref{Gamma_definitions}).  Then,
\begin{equation}
\begin{aligned}
    \Gamma^>_\mathbf{p}=&\; -\frac{1}{2p^0}{\rm Tr} \left[\sum_{i'=a,b,c,d} \slashed{p} \Pi^>_{(i')}(p)\right] \\
    =&\; -  \frac{G_F^2}{2p^0} \int \frac{\dd ^3\ensuremath{\boldsymbol\ell} \dd ^3 \mathbf{q} \dd ^3 \mathbf{u}}{(2\pi)^9 8 E_\ell E_q E_u}(2\pi)^4 \delta^{(3)}(\mathbf{u }-  
    \mathbf{q} - \boldsymbol\ell + \mathbf{p} )  \\
      & \qquad \times \Big[ (1-f_\ell) f_u (1-f_q) \delta(E_u-E_q-E_\ell + p^0){\rm Tr}^{+++}  \\
      & \qquad \qquad +(1- f_\ell)  (1- f_u)  f_q \delta(-E_u+E_q-E_\ell + p^0){\rm Tr}^{+--}  \\
      & \qquad \qquad +f_\ell (1- f_u) (1-f_q) \delta(-E_u-E_q+E_\ell + p^0){\rm Tr}^{--+}\Big],
      \label{Gamma_morethan}
\end{aligned}
\end{equation}
which together with $\Gamma_{\mathbf{p}}^{<}$ can be used to construct  a collision term for $f_{p}$,
\begin{equation}
C[f_{p}] \equiv (1-f_p)\Gamma^<_\mathbf{p}-f_p \Gamma^>_\mathbf{p},
\end{equation}
based on the generalised Boltzmann equation~(\ref{eq:gainlossboltzmann}).

In order to match the existing results in the literature, e.g, equation~(8) and table~I of~\cite{Grohs2016},  we change the 4-momentum variables to $P_{i} = (E_{i},\mathbf{p}_{i})$, $i=1,\ldots,4$.  Then, for $f_{1} \equiv f_{p}$, we find the collision term
\begin{align}
    C[f_1]=& \frac{1}{2E_1} \int \frac{\dd ^3 \mathbf{p}_{2} \dd ^3\mathbf{p}_{3}\dd ^3 \mathbf{p}_{4}}{(2\pi)^9 8 E_2 E_3 E_4}(2\pi)^4 \delta^{(4)}(P_1 +P_2-P_3-P_4 )F_r S_r\langle|\mathcal{M}_{r}|^2\rangle,
    \label{Boltzmann_representation}
\end{align}
where $F_r = (1-f_1)(1-f_2)f_3f_4 - f_1f_2(1-f_3)(1-f_4)$ is a phase space factor, and the symmeterised squared matrix elements
\begin{align}
&S_r\langle|\mathcal{M}_{r}|^2\rangle=S_1 \langle|\mathcal{M}_{\scaleto{\nu_e+e^-\leftrightarrow e^-+\nu_e}{7.5pt}}|^2\rangle+S_2 \langle|\mathcal{M}_{\scaleto{\nu_e+e^+\leftrightarrow e^++\nu_e}{7.5pt}}|^2\rangle+S_3 \langle|\mathcal{M}_{\scaleto{\nu_e+\bar{\nu}_e\leftrightarrow e^-+e^+}{7.5pt}}|^2\rangle
\end{align}
can be mapped to the traces of equations~(\ref{Gamma_lessthan}) and~(\ref{Gamma_morethan}) via
\begin{equation}
\begin{aligned}
    G_F^{-2} S_1 \langle|\mathcal{M}_{\nu_e+e^-\leftrightarrow e^-+\nu_e}|^2\rangle &\equiv - \left. {\rm Tr}^{+--} \right|_{P_2=(E_q,- \mathbf{q}),P_3=(E_u, - \mathbf{u}),
    P_4=(E_\ell, \ensuremath{\boldsymbol\ell})}, \\
    G_F^{-2} S_2 \langle|\mathcal{M}_{\nu_e+e^+\leftrightarrow e^++\nu_e}|^2\rangle &\equiv  -
    \left.  {\rm Tr}^{+++}\right|_{P_2=(E_u, \mathbf{u}),P_3=(E_q,\mathbf{q}), P_4=(E_\ell,  \ensuremath{\boldsymbol\ell})}, \\
  G_F^{-2} S_3 \langle|\mathcal{M}_{\nu_e+\bar{\nu}_e\leftrightarrow e^-+e^+}|^2\rangle &\equiv  -
   \left. {\rm Tr}^{--+}\right|_{P_2=(E_\ell,- \ensuremath{\boldsymbol\ell}),P_3=(E_q,\mathbf{q}),P_4=(E_u,-\mathbf{u})}.
\end{aligned}
\end{equation}
Thus, we have demonstrated that the neutrino damping rate calculated from the two-loop retarded self-energy at the (quasi)particle pole is indeed equivalent at leading order to the the usual Boltzmann collision term for $2 \to 2$ scattering from kinetic theory.


\section{Leading-order calculation of the damping rate}\label{weak_rate_calculation}

We show in this section how to reduce the self-energy expressions~(\ref{Kite1}) to~(\ref{Bubble2}) to the leading-order mode-dependent interaction rate~(\ref{weak_rate}).
Plugging into equations~(\ref{Kite1}) to~(\ref{Bubble2})  the free-fermion Wightman propagators~(\ref{eq:freefermionwightman}) and the traces~(\ref{eq:tra}) to~(\ref{eq:trd}) with $u = \ell-p+q$,  the self-energy contributions can be written as
\begin{equation}
\begin{aligned}
\label{general_selfenergy_integral}
{\rm Tr} \left[ \slashed{p} \Pi^<_{(i)}(p)\right] =  & \frac{\mathcal{C}_{i} \; G_F^2}{(2\pi)^5} \int \dd^4 \ell \dd^4q \;  \mathcal{F}(p^{0}, q^0,\ell^0)  \delta (\ell^2) \delta (q^2-m_{e}^2) \delta((\ell-p+q)^2-m_{e}^2)  \\
& \times \Big[ A_{(p\ell)}^i \;  m_{e}^2 (p\cdot \ell)  + A^i_{(pq)(\ell p)} (p\cdot q)(\ell \cdot p) +  A_{(pq)^2}^i (p\cdot q)^2 + A_{(p\ell)^2}^i(p\cdot \ell )^2\Big],
\end{aligned}
\end{equation}
where
\begin{equation}
    \mathcal{F}(p^0,q^0,\ell^0)\equiv \Big[f_D(|\ell^0 + q^0 -p^0|)-\theta (\ell^0 + q^0 -p^0)  \Big]\Big[f_D(|\ell^0|)-\theta (- \ell^0)  \Big]\Big[f_D(|q^0|)-\theta (-q^0)  \Big],
\end{equation}
and the various coefficients (${\cal C}_{i}$, $A^{i}_{(p \ell)}$, etc.) are given in table~\ref{tab:coeff1}.  

\begin{table}[t]
\centering
\begin{tabular}{|c||c|c|c|}
\hline
& $i=a,b$ & $i=c$ & $i=d$ \\
\hline
\hline
${\cal C}_{i}$ & 64 & $-128$ & $-128$ \\
\hline
$A^{i}_{(p\ell)}$ &   $g_{V,e}-g_{A,e}$ &$ -(g_{V,e}^2-g_{A,e}^2)$ & 0\\
\hline
$A^{i}_{(pq)(\ell p)}$ &$-4(g_{V,e}+g_{A,e})$ & $2 (g_{V,e}+g_{A,e})^2$ & 2\\
\hline
$A^{i}_{(pq)^{2}}$ & $-2(g_{V,e}+g_{A,e})$ &$ 2(g_{V,e}^2 + g_{A,e}^2)$ & 1\\
\hline
$A^{i}_{(p\ell)^{2}}$ & $ -2(g_{V,e}+g_{A,e})$ & $ (g_{V,e}+g_{A,e})^2$ & 1\\
\hline
\end{tabular}
\caption{Coefficients appearing in the weak-rate and self-energy integrals~(\ref{weak_rate}), (\ref{general_selfenergy_integral}), and (\ref{eq:selfself}).\label{tab:coeff1}}
\end{table}

To simplify equation~\eqref{general_selfenergy_integral}, observe that it is of the form
\begin{equation}
\begin{aligned}
   I & = \int \dd^4 q \dd^4 \ell \; f(q,\ell,p)  \;  \delta(\ell^2)\delta(q^2-m_{e}^2)\delta((\ell + q - p)^2 -m_{e}^2) \\
   &=   \int \frac{ \dd^3 \boldsymbol{\ell} \dd^3 \mathbf{q}}{4 |\boldsymbol{\ell}| E_q}  \left. \sum_{\epsilon,\tau = \pm 1} f(q,\ell,p)  \;\delta((\ell + q - p)^2 -m_{e}^2)\right|_{\overset{ q^0=\epsilon E_q}{\ell^0=\tau |\boldsymbol{\ell}}|},
   \label{eq:integral0}
\end{aligned}
\end{equation}
where $E_q^2 = |\mathbf{q}|^2 + m_{e}^2$, $f(q,\ell,p)$ is a scalar function independent of the Lorentz contraction $\ell \cdot q$, and the second equality follows from the fact that the first two Dirac deltas in $\ell^{2}$ and $q^{2}$ simply put the corresponding particles on their mass shells.  To evaluate the remaining Dirac delta distribution, we first parameterise the 3-momenta in spherical coordinates,
\begin{equation}
\begin{aligned}
    \mathbf{p} & =    |\mathbf{p}|(0,0,1)^T, \\
    \boldsymbol{\ell} & = |\boldsymbol{\ell}| (0,\sin \alpha ,\cos \alpha)^T, \\
    \mathbf{q} & =    |\mathbf{q}| (\sin\theta\sin\beta,\sin\theta \cos\beta,\cos\theta)^T, 
\end{aligned}
\end{equation}
so that the integral~(\ref{eq:integral0}) becomes
\begin{equation}
\begin{aligned}
    I=&  \frac{2\pi}{4} \iint_0^\infty\dd |\boldsymbol{\ell}|   \dd |\mathbf{q}|  \int_0^{2\pi}\dd\beta\int_{-1}^{+1}\dd\cos\alpha \; \dd\cos\theta \\
  &\qquad \qquad  \times \frac{|\mathbf{q}|^2 |\boldsymbol{\ell}|}{E_q} \left. \sum_{\epsilon,\tau = \pm 1} f(q,\ell,p)  \;\delta((\ell + q - p)^2 -m_{e}^2)\right|_{\overset{ q^0=\epsilon E_q}{\ell^0=\tau |\boldsymbol{\ell}|}}.
  \label{eq:integral1}
\end{aligned}
\end{equation}
Similarly, the  Lorentz contractions can now be written as
\begin{equation}
\begin{aligned}
    &\ell \cdot p= \tau |\boldsymbol{\ell}| |\mathbf{p}| - |\mathbf{p}| |\boldsymbol{\ell}| \cos \alpha,\\
    &p \cdot q = \epsilon E_q  |\mathbf{p}| - |\mathbf{p}| |\mathbf{q}| \cos \theta,\\
        &\ell \cdot q = \epsilon \tau E_q  |\boldsymbol{\ell}| - |\boldsymbol{\ell}| |\mathbf{q}| (\sin \alpha \sin \theta \cos \beta + \cos \alpha \cos \theta),
\end{aligned}
\end{equation}
from which we immediately deduce that the integrand~(\ref{eq:integral1}) depends on $\beta$ only through the Dirac delta distribution $\delta((\ell + q - p)^2 -m_{e}^2)$.

Then, following~\cite{Oldengott:2014qra}, we can solve the $\beta$-integral in~(\ref{eq:integral1}) by first identifying $\delta((\ell + q - p)^2 -m_{e}^2)\equiv \delta(g(\beta))$, which can be further decomposed to
\begin{equation}
\label{eq:deltaofg}
    \delta (g(\beta)) = \sum_{i} \frac{1}{|g'(\beta_i)|}\delta(\beta-\beta_i) =  \frac{\delta(\beta-\beta_1) + \delta(\beta-\beta_2)}{2 |\boldsymbol{\ell}| |\mathbf{q}| |\sin\theta \sin\alpha \sin \beta_0|}.
\end{equation}  
Here, $\beta_i$ denotes the roots of the function $g(\beta)$: in this case, there are two, $\beta_{1}$ and $\beta_{2}=2 \pi-\beta_{1}$, on the interval $\beta \in [0, 2\pi]$, given by
\begin{equation}
 \cos \beta_{i} = \frac{|\mathbf{q}| |\mathbf{p}| \cos\theta - \epsilon E_q |\mathbf{p}| + |\boldsymbol{\ell}| |\mathbf{p}| \cos \alpha - \tau |\boldsymbol{\ell}| |\boldsymbol{\ell}| + \epsilon \tau |\boldsymbol{\ell}| E_q - |\boldsymbol{\ell}| |\mathbf{q}| \cos \alpha \cos \theta }{|\boldsymbol{\ell}| |\mathbf{q}| \sin \theta\sin \alpha }\label{vanishing_cosine},
\end{equation}
and $|\sin \beta_{1}| = |\sin \beta_{2}| \equiv |\sin \beta_{0}|$; the derivative is $g'({\beta_{i}}) \equiv \left. \partial_{\beta} g \right|_{\beta_{i}} = -2 |\boldsymbol{\ell}| |\mathbf{q}| \sin \theta\sin\alpha \sin \beta_{i}$.
Then, substituting equation~(\ref{eq:deltaofg})  into equation~(\ref{eq:integral1}), yields
\begin{equation}
\label{eq:integral2}
I= \frac{2\pi}{4}\iint_0^\infty  \dd |\boldsymbol{\ell}| \dd |\mathbf{q}| \int_{-1}^{+1}\dd\cos\alpha \; \dd\cos\theta  \frac{|\mathbf{q}|^2 |\boldsymbol{\ell}|}{ E_q} \left. \sum_{\epsilon,\tau = \pm 1}\frac{f(q,\ell,p) \; \theta(1-\cos^2\beta_0)}{  |\boldsymbol{\ell}| |\mathbf{q}| |\sin\theta\sin\alpha \sin \beta_0|}\right|_{\overset{ q^0=\epsilon E_q}{\ell^0=\tau |\boldsymbol{\ell}|}},
\end{equation}
where we have inserted a Heaviside step function  $\theta(1-\cos^2\beta_0)$ to ensure that the condition $\cos^{2} \beta_{0} \leq 1$ is respected.

The next step is to recognise that, in equation~(\ref{eq:integral2}), the following two expressions are functionally equivalent:
\begin{equation}
    \frac{\theta(1-\cos^2\beta_0)}{ |\boldsymbol{\ell}| |\mathbf{q}|  |\sin\theta\sin\alpha \sin \beta_0|}= \frac{\theta(|\boldsymbol{\ell}|^2 |\mathbf{q}|^2 \sin^2 \theta\sin^2\alpha \; (1-\cos^2\beta_0))}{ \sqrt{|\boldsymbol{\ell}|^2 |\mathbf{q}|^2  \sin^2\theta\sin^2\alpha\; (1-\cos^2 \beta_0)}},
\end{equation}
where the r.h.s.~expression has the desirable property that the Heaviside step function and the square root take the same argument that is quadratic in $z \equiv \cos \theta$, i.e.,
\begin{equation}
\label{eq:quadratic}
    |\boldsymbol{\ell}|^2 |\mathbf{q}|^2  (1-z^{2})\sin^2\alpha\; (1-\cos^2 \beta_0)= \tilde{a} z^{2} + \tilde{b} z + \tilde{c},
\end{equation}
with coefficients
\begin{align}
    \tilde{a} = & -|\mathbf{q}|^2|\boldsymbol{\ell}-\mathbf{p}|^2 \leq 0, \label{eq:a} \\
    \tilde{b} = & -2\Big[-|\boldsymbol{\ell}|^2 |\mathbf{p}| |\mathbf{q}| \cos^2\alpha + (|\boldsymbol{\ell}| |\mathbf{p}|^2 |\mathbf{q}| + \epsilon E_q |\boldsymbol{\ell}| |\mathbf{p}| |\mathbf{q}| + \tau |\boldsymbol{\ell}|^2 |\mathbf{p}| |\mathbf{q}| - \epsilon \tau E_q |\boldsymbol{\ell}|^2 |\mathbf{q}| )\cos\alpha \nonumber\\
    & \qquad \qquad -\epsilon E_q |\mathbf{p}|^2 |\mathbf{q}| - \tau |\boldsymbol{\ell}| |\mathbf{p}|^2 |\mathbf{q}| + \epsilon\tau E_q |\boldsymbol{\ell}| |\mathbf{p}| |\mathbf{q}| \Big],  \label{eq:b}\\
    \tilde{c} = & - |\boldsymbol{\ell}|^2(|\mathbf{p}|^2+|\mathbf{q}|^2)\cos^2\alpha + 2(\epsilon E_q |\boldsymbol{\ell}| |\mathbf{p}|^2 + \tau |\boldsymbol{\ell}|^2 |\mathbf{p}|^2 -\epsilon \tau E_q |\boldsymbol{\ell}|^2 |\mathbf{p}|)\cos\alpha \nonumber \\
    & + |\boldsymbol{\ell}|^2(|\mathbf{q}|^2 - |\mathbf{p}|^2) - E_q^2(|\boldsymbol{\ell}|^2 + |\mathbf{p}|^2) + 2\epsilon  E_q |\boldsymbol{\ell}|^2 |\mathbf{p}| + 2 \tau E_q^2 |\boldsymbol{\ell}| |\mathbf{p}| -2 \epsilon \tau E_q |\boldsymbol{\ell}| |\mathbf{p}|^2, \label{eq:c}
\end{align}
and we note that $\tilde{a}$ is always negative or zero.

Because $\tilde{a} \leq 0$, the quadratic~(\ref{eq:quadratic}) represents a downward parabola with two real and non-degenerate roots $z_{\pm}$,
\begin{equation}
\label{eq:roots}
     z_\pm = \frac{\tilde{b}}{2 |\tilde{a}|} \pm \sqrt{\left(\frac{\tilde{b}}{2\tilde{a}}\right)^2 + \frac{\tilde{c}}{|\tilde{a}|}},
\end{equation} 
whenever $\tilde{b}^2 - 4 \tilde{a} \tilde{c} > 0$ is satisfied.  This also means that the Heaviside step function is  nonzero only in the region between non-degenerate $z_{\pm}$.
Then, rewriting the quadratic~(\ref{eq:quadratic}) in terms of its roots~(\ref{eq:roots}),
equation~(\ref{eq:integral2}) can now be recast as
\begin{equation}
\label{eq:integral3}
I= \frac{2\pi}{4}\iint_0^\infty  \dd |\boldsymbol{\ell}| \dd |\mathbf{q}| \frac{|\mathbf{q}|^2 |\boldsymbol{\ell}|}{  E_q} \! \! \int_{-1}^{+1}\dd\cos\alpha\frac{ \theta(\tilde{b}^2-4\tilde{a} \tilde{c})}{\sqrt{|\tilde{a}|}} \int_{z_-}^{z_+} \dd z \!\!\left.  \sum_{\epsilon,\tau = \pm 1} \frac{f(q,\ell,p) }{ \sqrt{(z-z_-)(z_+-z)}}\right|_{\overset{ q^0=\epsilon E_q}{\ell^0=\tau |\boldsymbol{\ell}|}}.
\end{equation}
To further reduce the number of integrals, we apply~(\ref{eq:integral3}) to the self-energy~\eqref{general_selfenergy_integral} to get
\begin{equation}
\begin{aligned}
{\rm Tr} \left[ \slashed{p} \Pi^<_{(i)}(p)\right]  
 =&  \frac{\mathcal{C}_i \; G_F^2}{4(2\pi)^4}\int_0^\infty \dd |\boldsymbol{\ell}| \dd|\mathbf{q}|  \frac{|\mathbf{q}|^2 |\boldsymbol{\ell}|}{ E_q} \int_{-1}^{+1}\dd\cos\alpha\sum_{\epsilon,\tau = \pm 1}\Bigg[\frac{ \theta(\tilde{b}^2-4\tilde{a} \tilde{c})}{\sqrt{|\tilde{a}|}}\mathcal{F}(p^0,q^0,\ell^0)\\
&   \times  \left( A^i_{(p\ell)} \;   I^{z}_{(p\ell)}  + A^i_{(pq)(\ell p)} I^z_{(pq)(\ell p)} + A^i_{(pq)^2} I^z_{(pq)^2} + A^i_{(p\ell)^2} I^z_{(pl)^2} \right) \Bigg]
\Bigg|_{\overset{q^0=\epsilon E_q}{\ell^0=\tau |\boldsymbol{\ell}|}},
\label{eq:selfself}
 \end{aligned}
 \end{equation}
where, for $(mn) = (p \ell), (p q)(\ell p), (p q)^{2}, (p \ell)^{2}$, we have defined
\begin{equation}
\begin{aligned}
\label{eq:izmn}
    I^z_{(mn)}& =\int_{z_-}^{z_+} \dd z\left(\frac{G^0_{(mn)} + z G^1_{(mn)} + z^2 G^2_{(mn)}}{\sqrt{(z-z_-)(z_+-z)}}\right) \\
&    = \pi\left[G^0_{(mn)} + \frac{\tilde{b}}{2|\tilde{a}|} G^1_{(mn)} + \left(\frac{3 \tilde{b}^2 + 4 \tilde{c} | \tilde{a} |}{8 \tilde{a}^2}\right) G^2_{(mn)}\right],
\end{aligned}
\end{equation}
with the $z$-independent coefficients  $G^{j}_{(mn)}$, $j=0,1,2$, given in table~\ref{tab:coeff2}.

\begin{table}[t]
\centering
\begin{tabular}{|c||c|c|c|}
\hline
& $j=0$ & $j=1$ & $j=2$ \\
\hline
\hline
$G^j_{(p \ell)}$ & $m_{e}^2(\tau |\boldsymbol{\ell}| |\mathbf{p}| - |\boldsymbol{\ell}| |\mathbf{p}| \cos\alpha)$ & 0 & 0 \\
\hline
$G^j_{(p q)(\ell p)}$ & $\epsilon E_q |\mathbf{p}| (\tau |\boldsymbol{\ell}| |\mathbf{p}| - |\boldsymbol{\ell}| |\mathbf{p}| \cos\alpha)$ & $|\mathbf{q}| |\mathbf{p}| (|\boldsymbol{\ell}| |\mathbf{p}| \cos \alpha - \tau |\boldsymbol{\ell}| |\mathbf{p}|)$ & 0 \\
\hline
$G^j_{(p q)^{2}}$ & $E_q^2 |\mathbf{p}|^2$ &  $-2 \epsilon E_q |\mathbf{p}|^2 |\mathbf{q}|$ & $|\mathbf{q}|^2|\mathbf{p}|^2$ \\
\hline
$G^j_{(p \ell)^{2}}$ & $(\tau |\boldsymbol{\ell}| |\mathbf{p}| - |\boldsymbol{\ell}| |\mathbf{p}| \cos\alpha)^2$ & 0 & 0 \\
\hline
\end{tabular}
\caption{Coefficients appearing in the weak-rate and self-energy integrals~(\ref{weak_rate}) and (\ref{eq:izmn}).\label{tab:coeff2}}
\end{table}

Putting it all back into equation~(\ref{eq:selfself}), and noting that the mode-dependent interaction rate, as defined in equation~(\ref{eq:master}), is equivalently
\begin{equation}
\Gamma_{\mathbf{p}} =\frac{(e^{p^0/T}+1)}{ 2p^0} \sum_i {\rm Tr} \left[ \slashed{p} \Pi^<_{(i)}(p)\right],
\end{equation}
we obtain the final result~(\ref{weak_rate}) for $p^{0} = |\mathbf{p}|$.

\bibliographystyle{utcaps}
\bibliography{bib}
\end{document}